\documentclass{emulateapj}

\usepackage{graphicx}
\usepackage{amsmath}
\usepackage{url}

\usepackage{apjfonts}
\usepackage{microtype}

\usepackage{float}
\usepackage{subcaption}

\renewcommand{\vec}{\mathbf}

\graphicspath{{./images/}}

\shorttitle{Generation of Magnetohydrodynamic Waves}
\shortauthors{S. J. Mumford et al.}

\begin{document}
\title{Generation of Magnetohydrodynamic Waves in Low Solar Atmospheric Flux Tubes by Photospheric Motions.}

\author{S. J. Mumford$^{1}$, V. Fedun$^{2, 1}$ and R. Erd\'elyi$^{1}$.}
\affil{$^{1}$Solar Physics \& Space Plasma Research Centre (SP$^{2}$RC), School of Mathematics and Statistics,\\ The University of Sheffield, Hicks Building, Hounsfield Road, Sheffield, S3 7RH U.K.}
\email{s.mumford@sheffield.ac.uk}
\affil{$^{2}$Solar Physics \& Space Plasma Research Centre (SP$^{2}$RC), Department of Automatic Control and Systems Engineering, The University of Sheffield, Mappin Street, Sheffield, S1 3JD, U.K.}

\begin{abstract}
Recent ground- and space-based observations reveal the presence of small-scale motions between convection cells in the solar photosphere.
In these regions small-scale magnetic flux tubes are generated due to the interaction of granulation motion and background magnetic field.
This paper studies the effects of these motions, on magnetohydrodynamic wave excitation from broadband photospheric drivers.
Numerical experiments of linear magnetohydrodynamic wave propagation in a magnetic flux tube embedded in a realistic gravitationally stratified solar atmosphere between the photosphere and the low choromosphere (above $\beta = 1$) are performed.
Horizontal and vertical velocity field drivers mimic granular buffeting and solar global oscillations.
A uniform torsional driver as well as Archimedean and logarithmic spiral drivers mimic observed torsional motions in the solar photosphere.
The results are analysed using a novel method for extracting the parallel, perpendicular and azimuthal components of the perturbations, which caters for both the linear and non-linear case.
Employing this method yields the identification of the wave modes excited in the numerical simulations and enables a comparison of excited modes via velocity perturbations and wave energy flux.
The wave energy flux distribution is calculated, to enable the quantification of the relative strengths of excited modes.
The torsional drivers excite primarily Alfv\'en modes ($\approx 60$\% of the total flux) with small contributions from the slow kink mode, and, for the logarithmic spiral driver, small amounts of slow sausage mode.
The horizontal and vertical drivers excite primarily slow kink or fast sausage modes respectively, with small variations dependent upon flux surface radius.
\end{abstract}

-

\section{Introduction}\label{sec:intro}
The solar atmosphere, stretching from the photosphere to the outer solar corona, is coupled by various magnetic structures.
 The lower layers of these structures are susceptible to perturbations in the photosphere caused by convective motions and by the solar global oscillations, which leads to disturbances propagating into the solar corona.
 In particular, these photospheric perturbations generate waves that propagate along the magnetic structures upwards through the chromosphere, transition region into the solar corona, transferring energy.

Such waves can be described by the magnetohydrodynamic (MHD) approximation which is shown to describe many modes and types of waves, in magnetic structures and their surrounding environment. 
 The study of wave propagation in the solar atmosphere is currently one of the major areas of research in solar physics. 
 It is widely accepted that MHD waves could provide one of the key ingredients in solving the atmospheric and coronal heating problem, however, an accurate mechanism for modelling the generation, transmission and release of this wave energy is yet to be fully developed.

In particular one of the main areas of study for the heating mechanism is the torsional Alfv\'en mode. 
This is an incompressible wave mode that, in a cylindrical geometry, naturally fitting a magnetic flux tube, can be described as a twisting perturbation of the magnetic field. 
For a review of current observations and theory regarding Alfv\'en waves in the solar atmosphere see \citet{mathioudakis2012}.

Modern high-resolution observations of the solar atmosphere, \textit{e.g.} \citet{bonet2008, bonet2010, wedemeyer-boehm2009, wedemeyer-boehm2012, wedemeyer2013, su2014}, have revealed vortex-type motions in different layers of the solar atmosphere and at various temporal and spatial scales.
 \citet{bonet2008} observed magnetic bright points (MBPs) spiralling into convective downdrafts.
 Such motions are a good candidate for exciting Alfv\'en waves in axially symmetric magnetic structures.
 These magnetic structures are likely to co-exist with the downdrafts in inter-granular lanes due to the increased magnetic field concentration in these regions \citep{shelyag2012}.
 \citet{bonet2010} expanded on this work by studying flows in the photosphere not directly linked to MBPs, while \citet{wedemeyer-boehm2009, wedemeyer-boehm2012} observe swirl motions in higher layers in the solar atmosphere, at larger spatial scales.

MHD wave propagation through the solar atmosphere is the subject of much computational study, with the primary goal of quantifying the free energy flux along the simulated magnetic structures.
 Most previous works have been devoted to the study of wave generation in two dimensional simulations \textit{e.g.}  \citet{bogdan2003, hasan2005, hasan2008, 2008SoPh..251..589K, fedun2011}, where horizontal and vertical drivers mimicking granular buffeting are used to excite oscillations. 
 However, three-dimensional simulations, have been employed and torsional motions are becoming the focus of numerical study, \textit{e.g.} \citet{2010HiA....15..354P, 2010ApJ...719..357F, fedun2011-a, shelyag2012a, vigeesh2012, 2013hsa7.conf..806C}. 
 In \citet{fedun2011-a} and \citet{vigeesh2012} torsional drivers have been used as they efficiently excite Alfv\'en waves in the flux tubes studied.
 \citet{vigeesh2012} analyses the energy flux carried by these torsional Alfv\'en waves in comparison to waves excited by a horizontal driver and finds that the torsional driver excites the most energy flux in the torsional component, as would be expected.

In the present work three different torsional motions are considered for the excitation of MHD waves in an open magnetic flux tube. 
 Two of the three types of these highly torsional drivers have a radial component to the velocity, the logarithmic and Archimedean spirals. 
 Logarithmic spiral motions were fitted to MBPs observed by \citet{bonet2008}, the Archimedean spiral is inspired by the spiral motion observed in \citet{wedemeyer-boehm2009, wedemeyer2013}. 
 As well as these, for comparison the already studied uniform torsional and horizontal and vertical motions are also implemented.

In this work we analyse the spectra of wave modes excited by different types of photospheric motions, keeping a fixed period and amplitude of all modelled drivers.
 The rest of this paper is organised thus: Sect. \ref{sec:simconfig} describes the configuration of the numerical domain and the photospheric drivers; Sect. \ref{sec:sim_and_analysis} is devoted to the analysis method and precedes to apply it to the results of the simulations, studying the phase speeds and energy flux of the excited wave modes; finally, we summarise the results and conclude in Sect. \ref{sec:conclusion}.

\section{Simulation Configuration} \label{sec:simconfig}
This work employs a 3D self-similar approach to construct the magnetic field in the form of an axial symmetric flux tube configuration similar to \citet{fedun2011, fedun2011-a, gent2013, gent2014} and based on \citet{schlueter1958, deinzer1965, schuessler2005}, providing a well-known and tested basis for the model.
 The footpoint of the magnetic flux tube is located at the centre of the domain in the horizontal plane at height $z = 0.0$ Mm, and has a field strength at the footpoint of $120$ mT ($1200$ Gauss). 
 This magnetic field strength corresponds to the typical value of magnetic field in a magnetic bright point \citep{shelyag2010, jess2010}.

Constructing a magnetic field using the self-similar approach generates a non-potential field. 
 Choosing a normalised exponential configuration for the expansion of the flux tube and the vertical decrease in magnetic field strength also ensures that $\nabla\cdot\vec{B} = 0 $ is analytically conserved by the model. The non-potential approach of this work differers from the potential-field configuration of \citet{vigeesh2012}.

\begin{figure}[h]
	\includegraphics[width=\columnwidth]{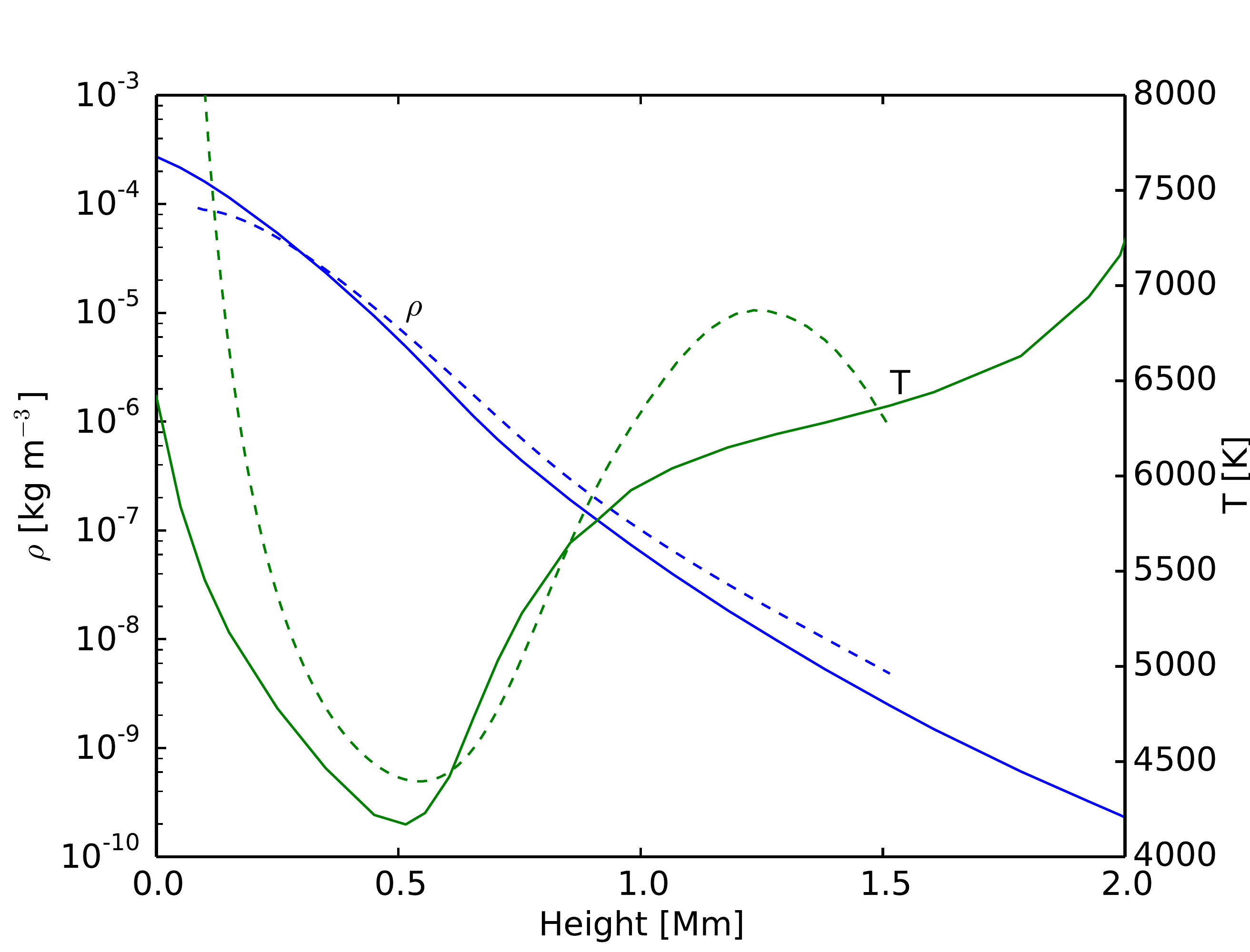}
	\caption{Temperature $T$ (green lines) and density $\rho$ (blue lines) from the quiet Sun VALIIIC \citep{Vernazza1981} atmosphere implemented as an equilibrium background for the numerical model (solid lines). The corresponding values at the centre of the computational domain (centre of the flux tube) are shown as dashed lines.}
	\label{fig_VALIIIC}
\end{figure}

The background non-magnetic atmosphere implemented here is based on the VALIIIC model of the quiet Sun atmosphere \citep{Vernazza1981} which is shown in Fig. \ref{fig_VALIIIC}.
 The changes in density and gas pressure due to the presence of the magnetic field are computed by employing magneto-hydrostatic equilibrium \textit{i.e.},
\begin{equation}
	(\mathbf{B}\cdot \nabla)\mathbf{B} + \nabla\left(\frac{\mathbf{B}^2}{2}\right) + \nabla p = \rho\mathbf{g}.
	\label{mhs-condition}
\end{equation}
This method has been used previously by \citet{shelyag2009} and \citet{fedun2011}. The computed background sound and Alfv\'{e}n speed profiles along the vertical axis of the magnetic flux tube and at the edge of the computational domain are shown in Fig. \ref{fig_VAL_va_cs}.
For a discussion of cutoff frequencies in this domain see the Appendix.

The vertical extent of the computational domain is chosen such that the $\beta =1$ region is avoided.
The reason for this choice is that this work is focusing on wave generation and excitation and quantifying the efficiency of different drivers, not on the mode conversion that happens around the $\beta = 1$ region.
Instead, we aim to investigate what distribution of wave modes are excited by the range of drivers present in the photosphere.
The models used for these drivers are inspired by observing the footpoint motion of flux tubes \textit{e.g.} \citet{bonet2008}.

\begin{figure}[h]
	\includegraphics[width=\columnwidth]{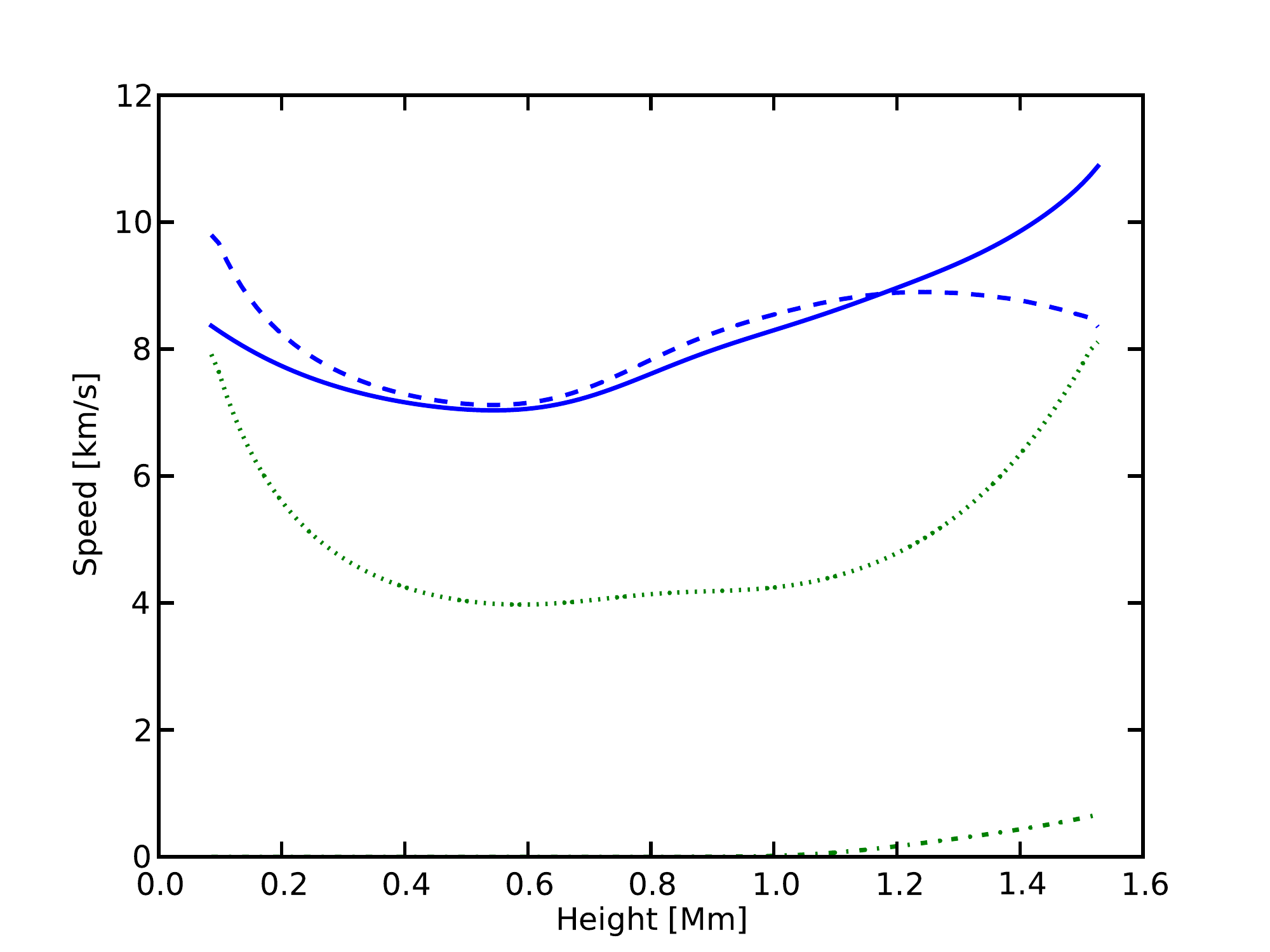}
	\caption{Sound $v_s$ and Alfv\'en $v_A$ speeds in the simulation domain, plotted along the axis of the magnetic flux tube, where the magnetic field is strong, and in the weakly magnetic background at the edge of the domain. The blue solid line is the sound speed $c_s$ in the background of the domain, the blue dashed line is the sound speed $c_s$ in the centre of the flux tube, the green dot-dashed line is the Alfv\'en speed $v_A$ in the background and the green dotted line is the Alfv\'en speed $v_A$ in the centre of the flux tube.}
	\label{fig_VAL_va_cs}
\end{figure}

The code used to carry out the simulations is the Sheffield Advanced Code (SAC) initially developed by \citet*{shelyag2008} and based on the Versatile Advection Code (VAC) \citep[see][for details]{toth1996}. 
 SAC is a fully non-linear MHD solver designed to solve various HD and MHD problems in non-magnetised and magnetised media, on top of any static background conditions including the strongly gravitationally stratified solar atmosphere employed in this work. 
 A schematic of the computational domain is represented in Fig. \ref{fig:domain_and_surface}. 
 The spatial extent of the domain in the $x$, $y$ and $z$ directions, respectively, are, $2.0 \times\ 2.0\ \times\ 1.6$ Mm$^3$, with an origin in the $z$ direction of $0.061$ Mm above the photosphere. The domain is divided up into $128^3$ grid points giving a physical size of $15.6\ \times\ 15.6\ \times\ 12.5$ km$^3$ for each grid cell. 
 Open boundary conditions are employed at all boundaries, which allow most linear perturbations to leave the domain without significant reflection.

\subsection{Footpoint Drivers}\label{sec:drivers}

 \begin{figure}[h]
 	\centering
 	\begin{subfigure}[b]{0.9\columnwidth}
 		\includegraphics[width=\columnwidth]{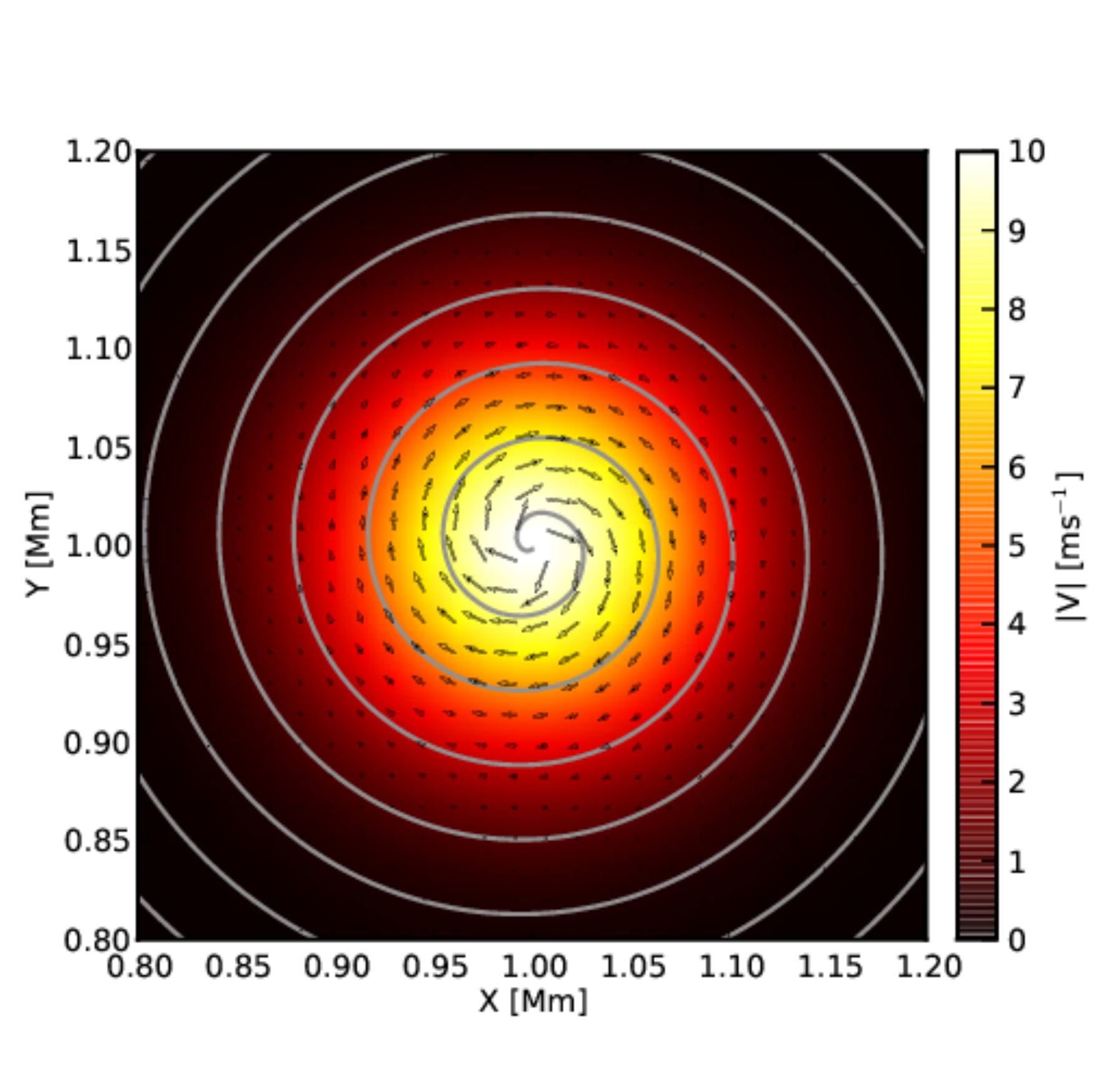}
 		\caption{Archimedean Spiral Driver.}
 	\end{subfigure}
 	
 	\begin{subfigure}[b]{0.9\columnwidth}
 		\includegraphics[width=\columnwidth]{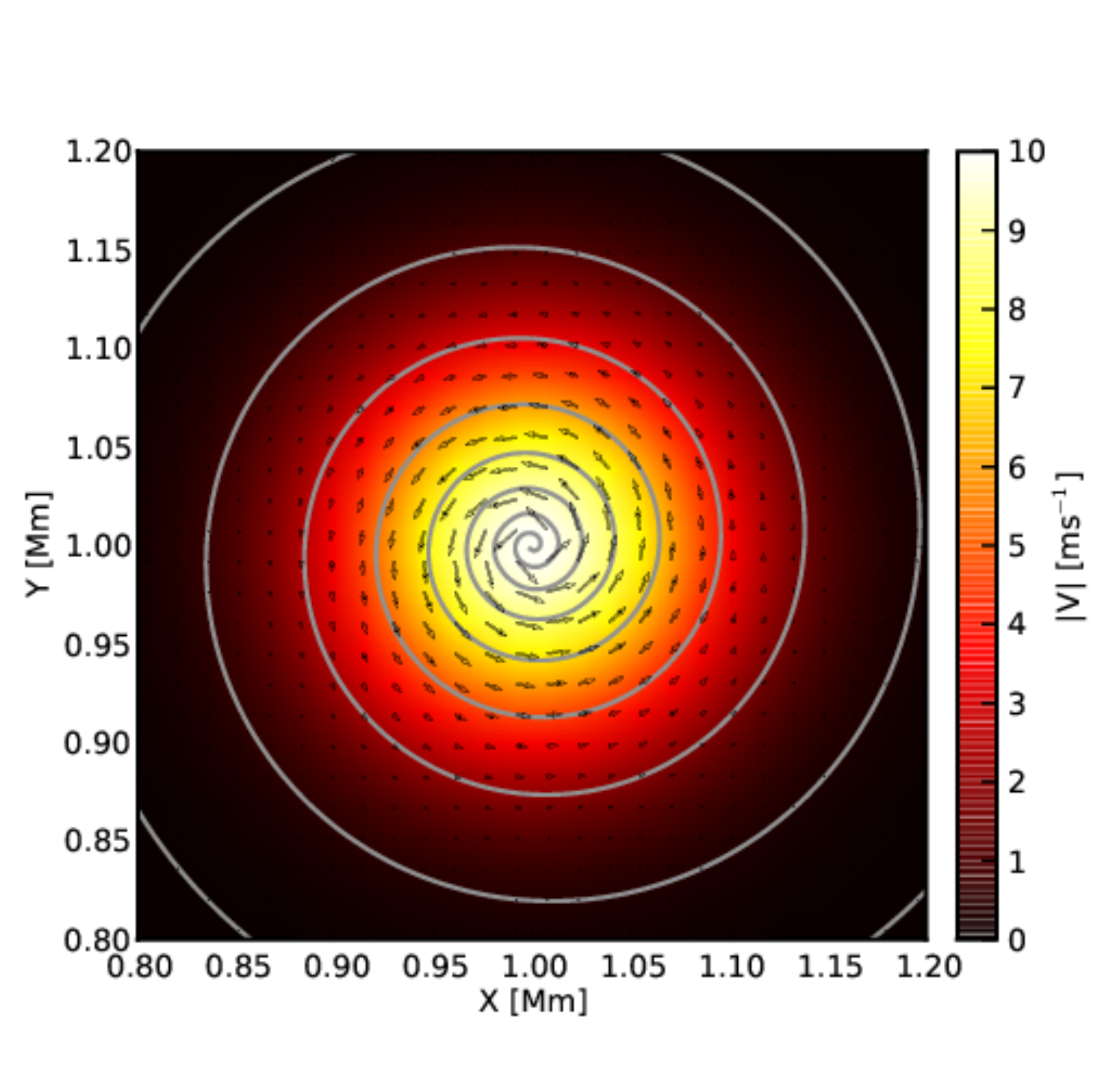}
 		\caption{Logarithmic Spiral Driver.}
 	\end{subfigure}
 	\caption{Horizontal cuts through the spiral driver at the peak amplitude height $z = 0.01$ Mm for the two spiral drivers. White lines are streamlines of the velocity vector field, sampled as black arrows, over plotted on the velocity magnitude $|V|$.}
 	\label{fig:spiral_driver_cut}
 \end{figure}

The equilibrium at the base of the simulated magnetic flux tube is perturbed to excite MHD waves propagating along the flux tube. 
 There are wealth of observed physical phenomena that could generate such motions on the Sun, for example plasma motion in the inter-granular lanes.
 Such as granular buffeting, the motion of the plasma at the top of the convection cells, which causes the plasma to be moved horizontally and vertically as the uprising granules, push the cooling plasma to the side. 
 These effects are modelled, for horizontal $(x)$ and vertical $(z)$ motions, in Equation \ref{eq:Vertical}:

\begin{equation}
	V_{x,z} = A \ e^{-\left(\frac{z^2}{\Delta z^2} + \frac{x^2}{\Delta x^2} + \frac{y^2}{\Delta y^2}\right)} \sin \left(2\pi \frac{t}{P}\right),
	\label{eq:Vertical}
\end{equation}
where $A$ is the amplitude, $\Delta x = \Delta y = 0.1$ Mm and $\Delta z = 0.05$ Mm. For all the drivers used in this work, the spatial distribution is a three-dimensional Gaussian profile, while the oscillation of the driver has an undamped sinusoidal profile of period $P=240$ s.

The uniform torsional driver is described by Equation \ref{eq:Suni}, which has a similar form to the horizontal and vertical drivers, with a spatial Gaussian shape.
 The uniform torsional driver is an adaptation from the simple vector field $F(x,y) = y\vec{x} - x\vec{y}$ scaled by $|\vec{r}|$, so that the field is of uniform intensity before the Gaussian profile is applied. 
 We model these motions as:
 
\begin{subequations}
	\begin{align}
		V_x &= A \frac{y}{\sqrt{x^2 + y^2}}\ e^{-\left(\frac{z^2}{\Delta z^2} + \frac{x^2}{\Delta x^2} + \frac{y^2}{\Delta y^2}\right)} \sin \left(2\pi \frac{t}{P}\right), \label{eq:Suni}\\
		V_y &= - A \frac{x}{\sqrt{x^2 + y^2}}\ e^{-\left(\frac{z^2}{\Delta z^2} + \frac{x^2}{\Delta x^2} + \frac{y^2}{\Delta y^2}\right)} \sin \left(2\pi \frac{t}{P}\right).
	\end{align}
\end{subequations}

As discussed in Sect. \ref{sec:intro}, recent observations have found spiralling motions in various layers of the solar atmosphere.
 \citet{bonet2008} fitted a logarithmic spiral to MBP motion in the inter-granular lanes and found a good fit.
 This type of motion is mimicked by the following equations:
 
\begin{subequations}
\begin{align}
	V_x &= A \frac{\cos(\theta + \phi)}{\sqrt{x^2 + y^2}}\ e^{-\left(\frac{z^2}{\Delta z^2} + \frac{x^2}{\Delta x^2} + \frac{y^2}{\Delta y^2}\right)} \sin \left(2\pi \frac{t}{P}\right),\\
	V_y &= - A \frac{\sin(\theta + \phi)}{\sqrt{x^2 + y^2}}\ e^{-\left(\frac{z^2}{\Delta z^2} + \frac{x^2}{\Delta x^2} + \frac{y^2}{\Delta y^2}\right)} \sin \left(2\pi \frac{t}{P}\right),\label{eq:Slog}\\
		&\text{where}\notag\\
		&\theta = tan^{-1}\left(\frac{y}{x}\right),\ \phi = tan^{-1}\left(\frac{1}{B_L}\right).\notag	
\end{align}
\end{subequations}
where $B_L = 0.05$ and is a dimensionless expansion parameter for the logarithmic spiral. The value of $B_L$ is chosen arbitrarily to fit a few spiral rotations within the driver volume see Fig. \ref{fig:spiral_driver_cut}.

\citet{wedemeyer-boehm2009} observed more circular trajectories as well an Archimedean-type spirals.
 This Archimedean spiral-type motion is implemented in our simulations (see Equations \ref{eq:Sarch} below) and is compared to the logarithmic, uniform torsional-type and horizontal- and vertical-type drivers. 
 The Archimedean spirals are modelled as:
 
\begin{subequations}
\begin{align}
	V_x &= A \frac{B_Ax}{x^2 + y^2} \frac{y}{\sqrt{x^2 + y^2}}\ e^{-\left(\frac{z^2}{\Delta z^2} + \frac{x^2}{\Delta x^2} + \frac{y^2}{\Delta y^2}\right)} \sin \left(2\pi \frac{t}{P}\right),\\
	V_y &= - A \frac{B_Ay}{x^2 + y^2} \frac{x}{\sqrt{x^2 + y^2}}\ e^{-\left(\frac{z^2}{\Delta z^2} + \frac{x^2}{\Delta x^2} + \frac{y^2}{\Delta y^2}\right)} \sin \left(2\pi \frac{t}{P}\right),
	\label{eq:Sarch}
\end{align}
\end{subequations}
$B_A = 0.005$ is similar in nature to $B_L$, \textit{i.e.} a dimensionless expansion parameter.
The amplitude $A$ of all the drivers is set to $10$ ms$^{-1}$ for all the simulations performed in this work.

\section{Simulations and Analysis}\label{sec:sim_and_analysis}

\begin{figure*}
	\centering
	\includegraphics[width=\textwidth]{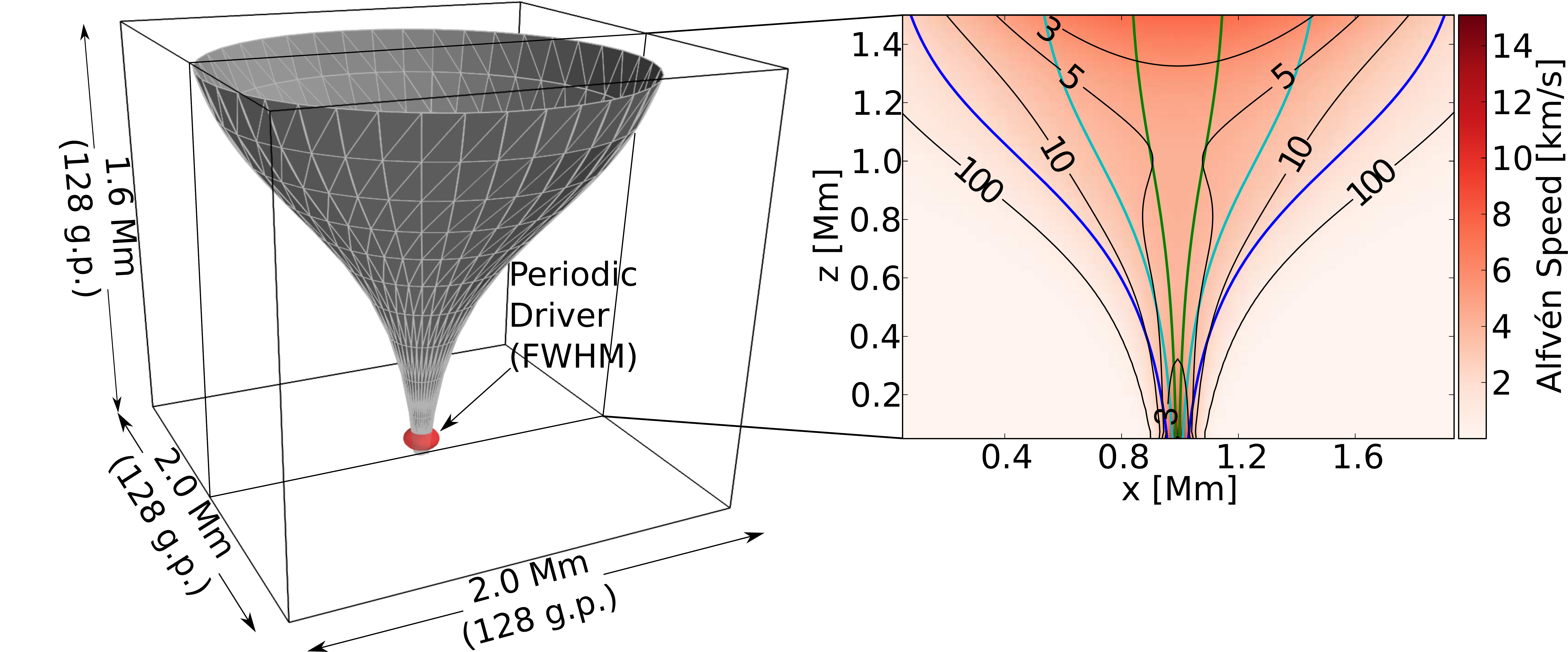}
	\caption{The left panel shows a schematic diagram of the computational domain. The volume of the periodic driver is highlighted at the base as FWHM of the Gaussian. The triangular polygon reconstruction of the surface from the field lines is shown, at low resolution for clarity. The right panel shows a 2D cut of the domain through the centre of the magnetic flux tube. Alfv\'en speed is plotted with plasma-$\beta$ contours over plotted (thin black lines). The coloured lines are magnetic field lines which form part of the surfaces of different radial distance from the unperturbed axis of the magnetic flux tube. The blue lines have a radius (at the top of the domain) of $r=936$ km, the cyan lines $r=468$ km and the green lines $r=156$ km.}
	\label{fig:domain_and_surface}
\end{figure*}

\subsection{Analysing MHD wave modes in three dimensions}\label{sec:3d_analysis}

The ideal MHD equations, in the case of linear perturbations in a uniform homogeneous magnetised plasma, have three independent eigenmodes, which correspond to the fast and slow magneto-acoustic and Alfv\'{e}n waves.
 These three modes have different properties, each with characteristics dependent upon the conditions of the plasma in which the wave is propagating. 
 Decomposing perturbations into these modes is not a trivial problem. 
 In the case of a gravitational stratification and presence of non-uniform magnetic field, the wave modes become coupled, and, therefore, the pure separation of these perturbations is not possible.
 The strength of this physical coupling depends on the ratio between the sound and Alfv\'en speeds, where this is large the coupling is weaker.
 Due to the changes in this ratio throughout the simulation domain, the coupling between the wave modes will be stronger in the centre of the domain where the Alfv\'en speed is higher.
 As well as this, depending on the geometry of the problem the wave modes can become degenerate in one or more spatial dimensions.
 In the case of cylindrical geometry this is the radial direction, which lead to the formation of the sausage, kink and fluting modes in thin flux tube theory.
 In our simulation domain this degeneration is present but in varying prevalence thought out the vertical extent of the domain, due to the expansion of the magnetic field.
 While these effects can make it harder to clearly identify the excited modes, in general these definitions from theory allow us to a good approximation quantify and describe the simulated waves.

MHD wave modes, where they have independent eigenmodes, can be totally isolated by their perturbations with respect to the magnetic field.
 In a uniform, high-$\beta$ plasma, the slow-magnetoacoustic mode perturbs the perpendicular component of velocity, while the fast-magnetoacoustic mode perturbs the velocity component in all directions.
 This decomposition becomes more complex under non-uniform conditions, where, as discussed both physical and geometrical considerations make the identification of the vector components with respect to the field and the interpretation of the results more difficult.

For two dimensional problems, it is trivial to compute a vector perpendicular to the magnetic field at any arbitrary point, which can then be attributed to the fast mode, and a vector parallel to the magnetic field which can be attributed to the slow mode.
 However, in a three-dimensional geometry, this definition breaks down as there are infinitely many vectors perpendicular to the magnetic field vector, and there is the additional presence of the Alfv\'en wave. 
 We employ a tool usually used by analytical models in cylindrical geometry, that of 'flux tubes'. 
 These are constructs where the tube contains a constant amount of magnetic flux, so a flux tube constructed vertically would have a constant amount of magnetic flux enclosed by its surface at height $h_1$ as $h_2$. 
 Flux tubes allow the identification of, in three-dimensional geometry, the fast, slow and Alfv\'{e}n modes, as the velocity perturbations parallel to the magnetic field and the flux tube; perpendicular to the flux tube and its surface and an azimuthal vector perpendicular to the magnetic field and parallel to the surface, respectively.

Numerically constructing a magnetic flux tube surface in a computational domain is a non-trivial problem.
 Defining an isosurface, \textit{i.e.} a surface along a suitable constant scalar value, to represent the flux tube surface is impractical due to the difficulty in finding such a constant quantity that is easily computable. 
 However, magnetic flux tubes can be constructed by using the properties of magnetic field lines. 
 A chosen area at some height in the domain can be used to trace the flux tube through all heights corresponding to that area. 
 The chosen area will enclose a constant value of magnetic flux, the integral of the field strength over the chosen area.
 If all the field lines crossing that area were selected and traced through the domain they would fill the entire volume of the corresponding flux tube. 
 To extract the surface of such a flux tube only the field lines intersecting the perimeter of the seed area are selected and traced.

Due to the axisymmetric nature of the magnetic field configuration implemented here, a circular seed area at the top of the numerical domain is chosen.
 Next, $100$ field lines from the edge are traced downwards through the domain. 
 These field lines are then fed into the vtkRuledSurfaceFilter in the VTK library \citep[www.vtk.org]{kitware2013} which generates surfaces comprised of a series of polygons from nearly parallel lines.
 This returns a number of triangular polygons from which a representation of the flux surface is obtained. 
 This flux surface is limited in accuracy by the angular resolution of the field line seeds and the resolution of the magnetic field line tracer. 
 For a schematic digram showing the construction of a low-resolution flux tube surface using 40 field lines, and 10\% of the field line tracer resolution, see the left panel in Fig. \ref{fig:domain_and_surface}.

Now having defined a set of polygons that lie on the flux surface, it is possible to construct a vector normal to the flux tube for each of the computed polygons.
 This can be achieved using the set of equations
\begin{equation}
	ax_{1,2,3}+by_{1,2,3}+cz_{1,2,3}+d=0,
\end{equation}
where the vector of the coefficients $a,b,c$ are then the vector normal to the polygon, $1, 2, 3$ are indicies for the three points that the plane is defined by and $x,y,z$ are the coordinates of the three vertices defining the polygon (a further check is performed in the calculations to ensure this vector is always radially away from the initial axis of symmetry of the domain).

The normal vector $(a,b,c)$ and the magnetic field vector can then be used to restrict the degrees of freedom given and decompose a new frame of reference for the velocity vector.
 As well as the already calculated vector perpendicular to the surface, the vector parallel to the magnetic field is computed as a unit vector in the direction of the magnetic field, the azimuthal unit vector is then $\vec{n_\phi}= \vec{n_\perp} \times \vec{n_\parallel}$ where $\vec{n_\perp},\ \vec{n_\parallel} $ are also both unit vectors. 
 The velocity vector components can then be computed by interpolating the velocity value to the coordinates of the surface polygons and projecting it onto the new coordinate system.

\subsection{Analysing Decomposed Velocity Perturbations}\label{sec:results}

\begin{figure}[h]
	\centering
	\begin{subfigure}[b]{0.9\columnwidth}
		\includegraphics[width=\columnwidth]{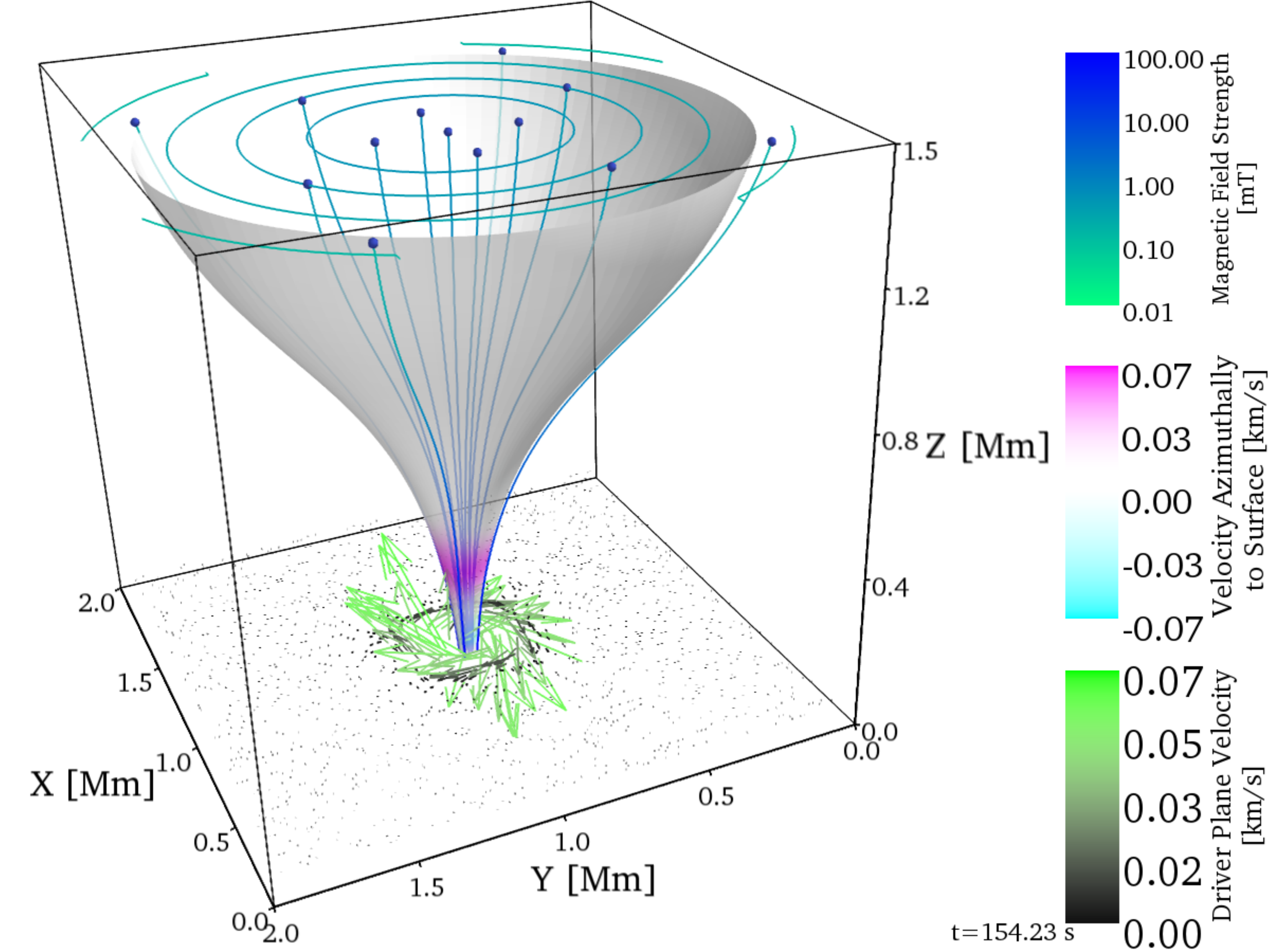}
		\caption{Snapshot at $t=154$ s}
	\end{subfigure}
	
	\begin{subfigure}[b]{0.9\columnwidth}
		\includegraphics[width=\columnwidth]{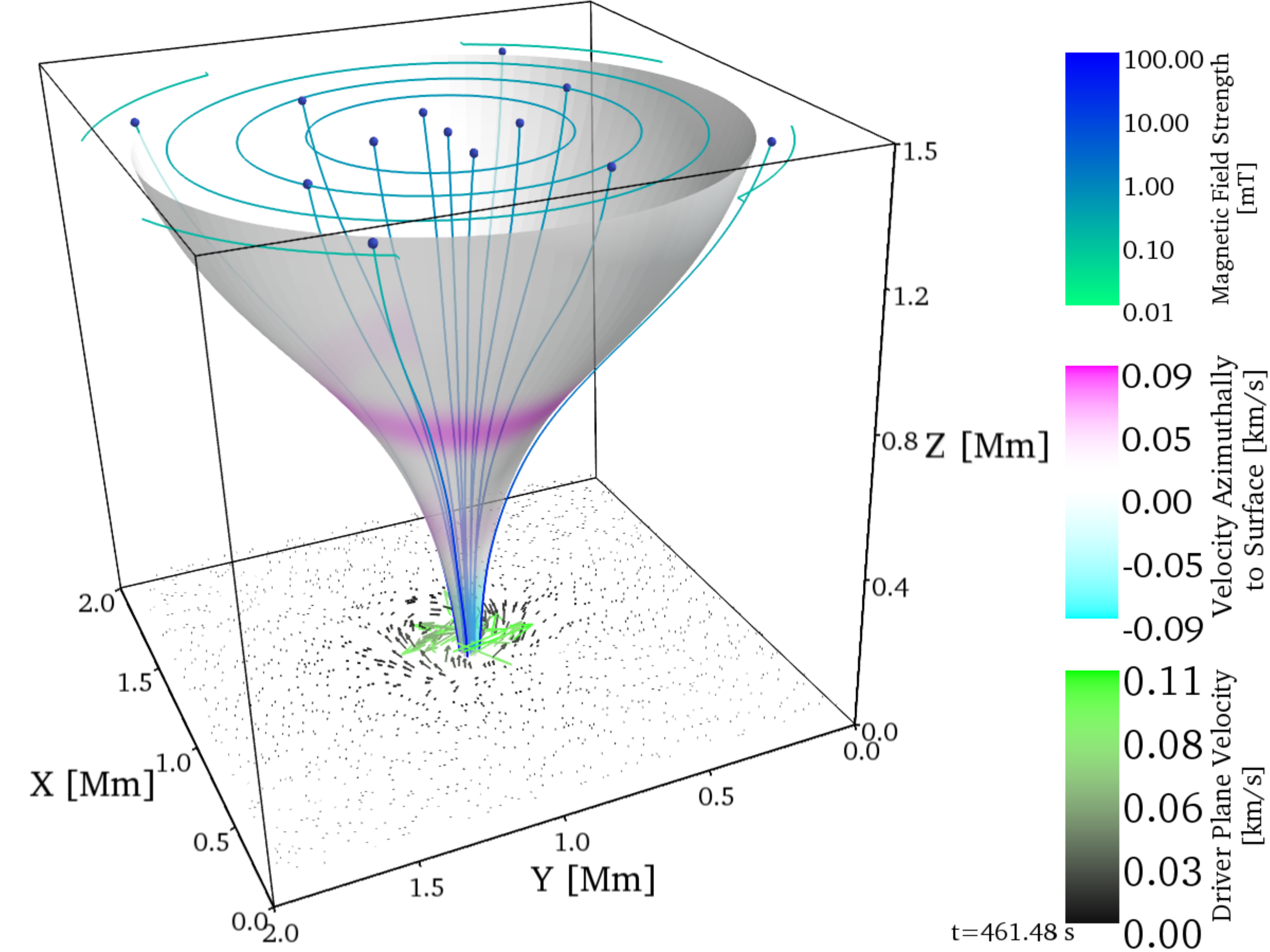}
		\caption{Snapshot at $t=461$ s}
	\end{subfigure}
	
	\begin{subfigure}[b]{0.9\columnwidth}
		\includegraphics[width=\columnwidth]{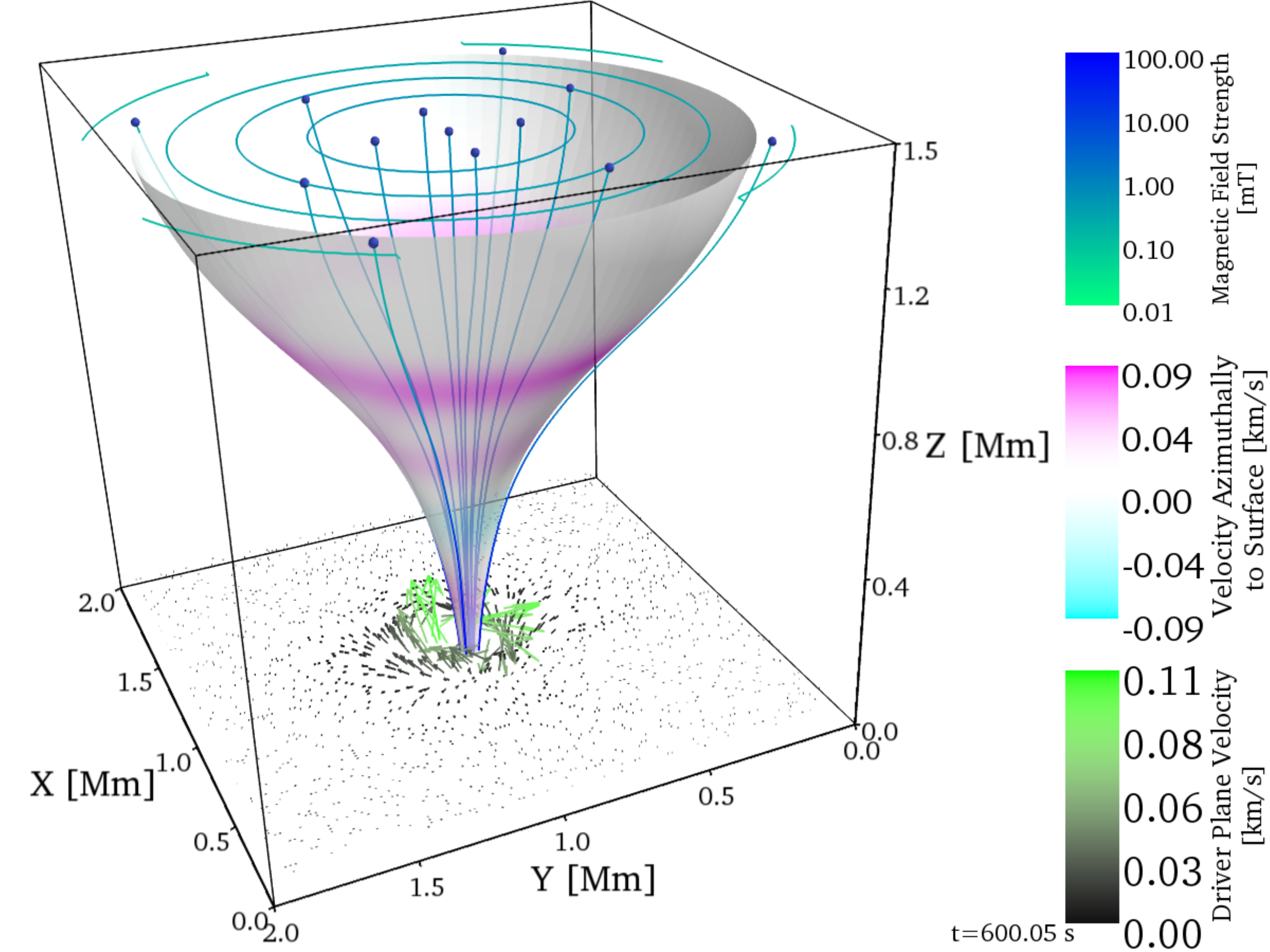}
		\caption{Snapshot at  $t=600$ s}
	\end{subfigure}
	\caption{Snapshots at three time steps of a 3D render of the simulation domain for the logarithmic spiral driver with a flux tube radius of $r = 936$ km (at the top of the computational domain). Shown in the domain are magnetic field lines and field strength contours in cyan, as well as the velocity vector field at the peak height of the driver shown as green and black arrows at the base, and the reconstructed surface coloured with the azimuthal velocity component ($V_\phi$).}
	\label{fig:frames_Suni_vphi}
\end{figure}

The results of applying the analysis which is discussed in Sect. \ref{sec:3d_analysis} are shown in Fig. \ref{fig:frames_Suni_vphi}, as snapshots at times $154$, $461$ and $600 \text{ s}$ of wave propagation along the magnetic flux tube surface as generated by a logarithmic spiral driver (see Equation \ref{eq:Slog}) with a period of $240$ s.
 In Fig. \ref{fig:frames_Suni_vphi} it can be seen that the torsional driver excites perturbations in each decomposed velocity component, \textit{e.g.} $V_\phi$,  $V_\perp$, $V_\parallel$, as would be expected for a driver that is not exactly a linear eigenmode of the system. 
 The strength and positions of these perturbations change as the simulations progress and the wave fronts travel along the tube. 
 Also shown is a vector plane at the peak vertical height of the driver, which illustrates the velocity field driving the oscillations.
To analyse the propagation of separate wave modes we extracted the velocity components along a magnetic field line on the flux surface and constructed time-distance diagrams for each component. 
 One magnetic field line is chosen at the beginning of the simulation and the values on the polygons between this field line and an adjacent field line are extracted for each time-step and presented in the time-distance diagrams in Fig. \ref{fig:All_TD_wave_30}. 
 The perturbations are assumed to be linear, as no correction is made for the (vertical) movement of the surface itself. 
 This assumption is verified by calculating the variation in the coordinates for the polygons at each time-step and it is found to be substantially less than one grid point for all the results presented here.

\subsection{Mode Identification}
\begin{figure*}
	\centering
	\begin{subfigure}[b]{0.49\textwidth}
		\includegraphics[width=\columnwidth]{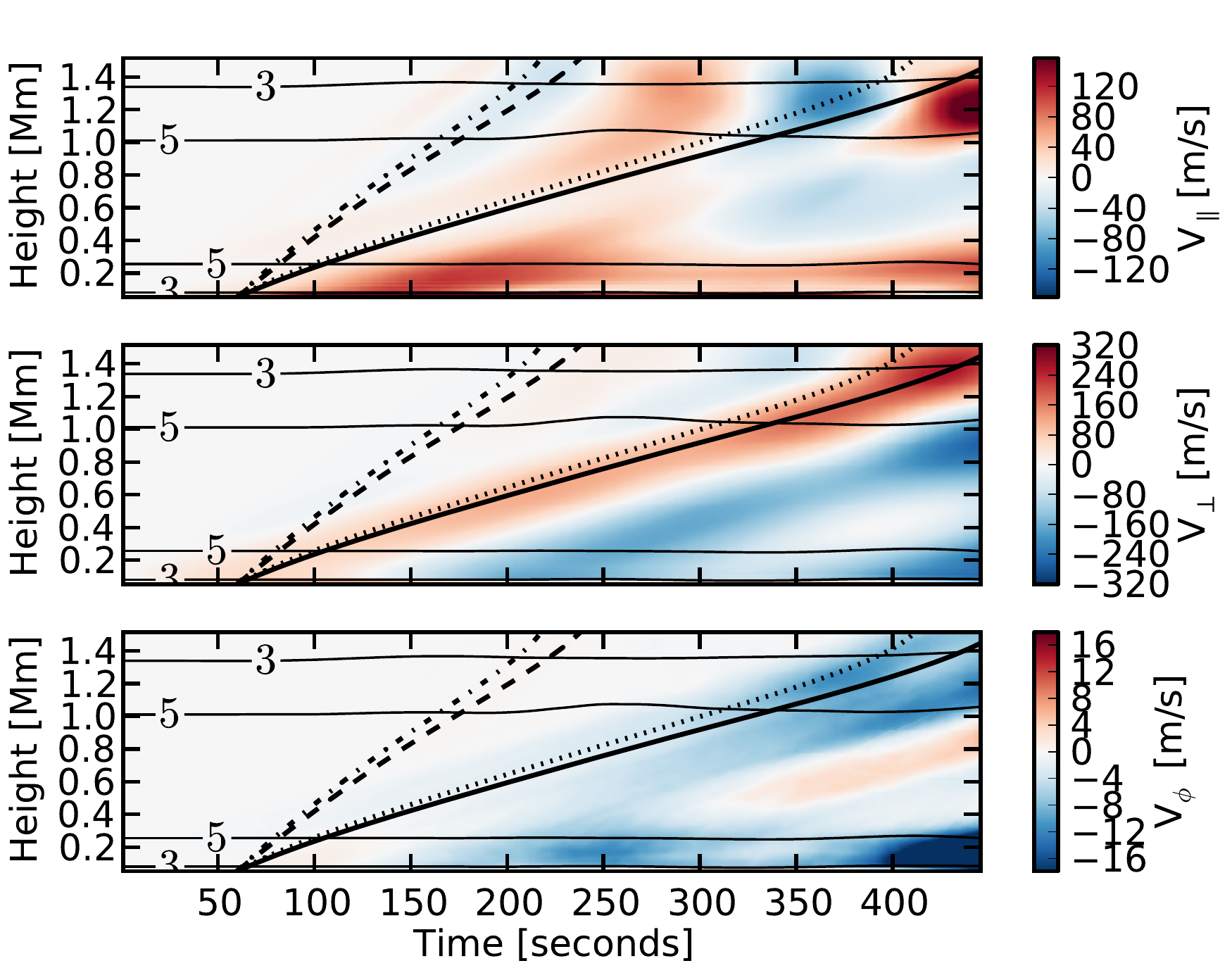}
		\caption{Horizontal Driver}
		\label{fig:All_TD_wave_30:horiz}
	\end{subfigure}
	\begin{subfigure}[b]{0.49\textwidth}
		\includegraphics[width=\columnwidth]{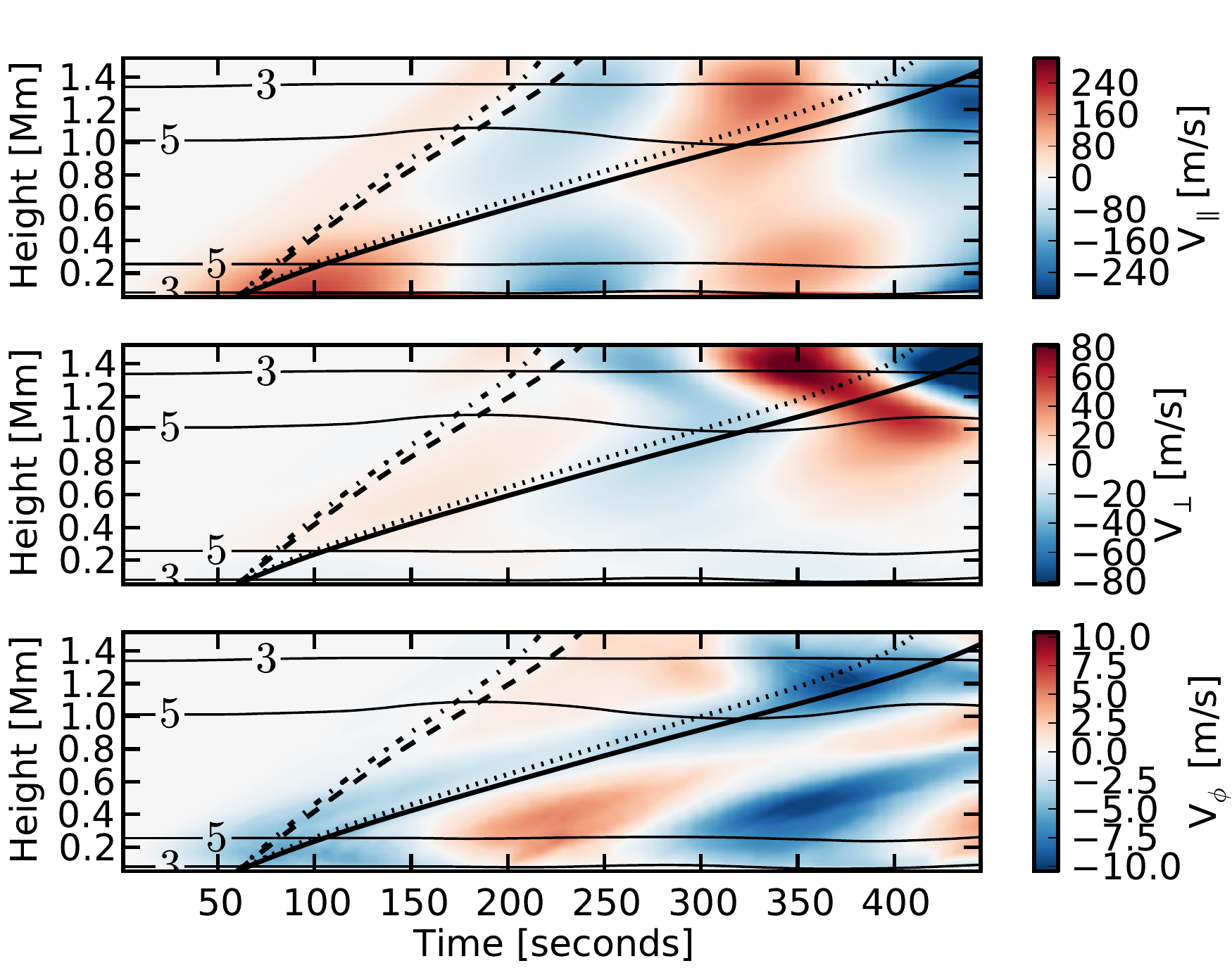}
		\caption{Vertical Driver}
		\label{fig:All_TD_wave_30:vert}
	\end{subfigure}
	
	\begin{subfigure}[b]{0.49\textwidth}
		\includegraphics[width=\columnwidth]{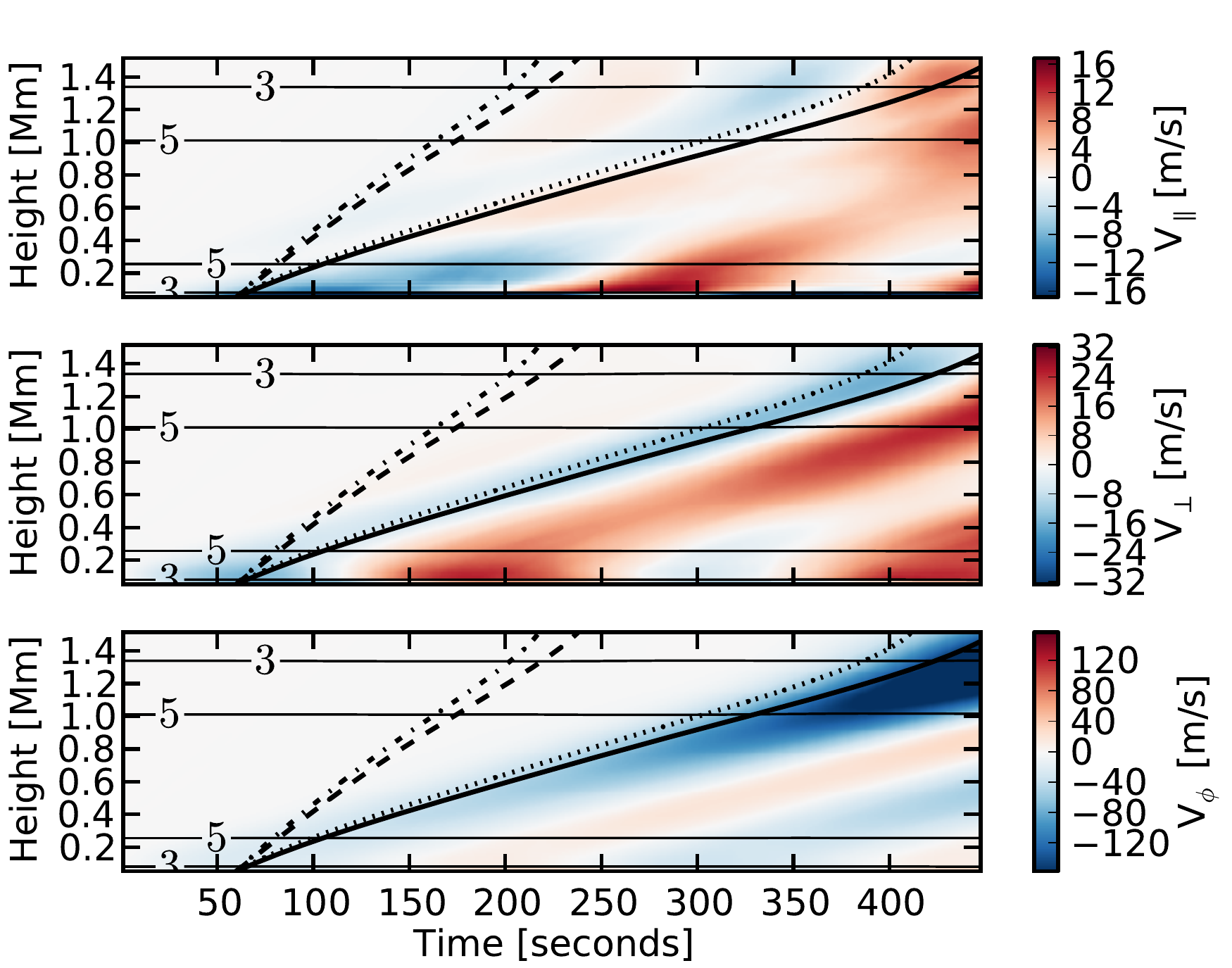}
		\caption{Uniform Torsional Driver}
		\label{fig:All_TD_wave_30:Suni}
	\end{subfigure}
	\begin{subfigure}[b]{0.49\textwidth}
		\includegraphics[width=\columnwidth]{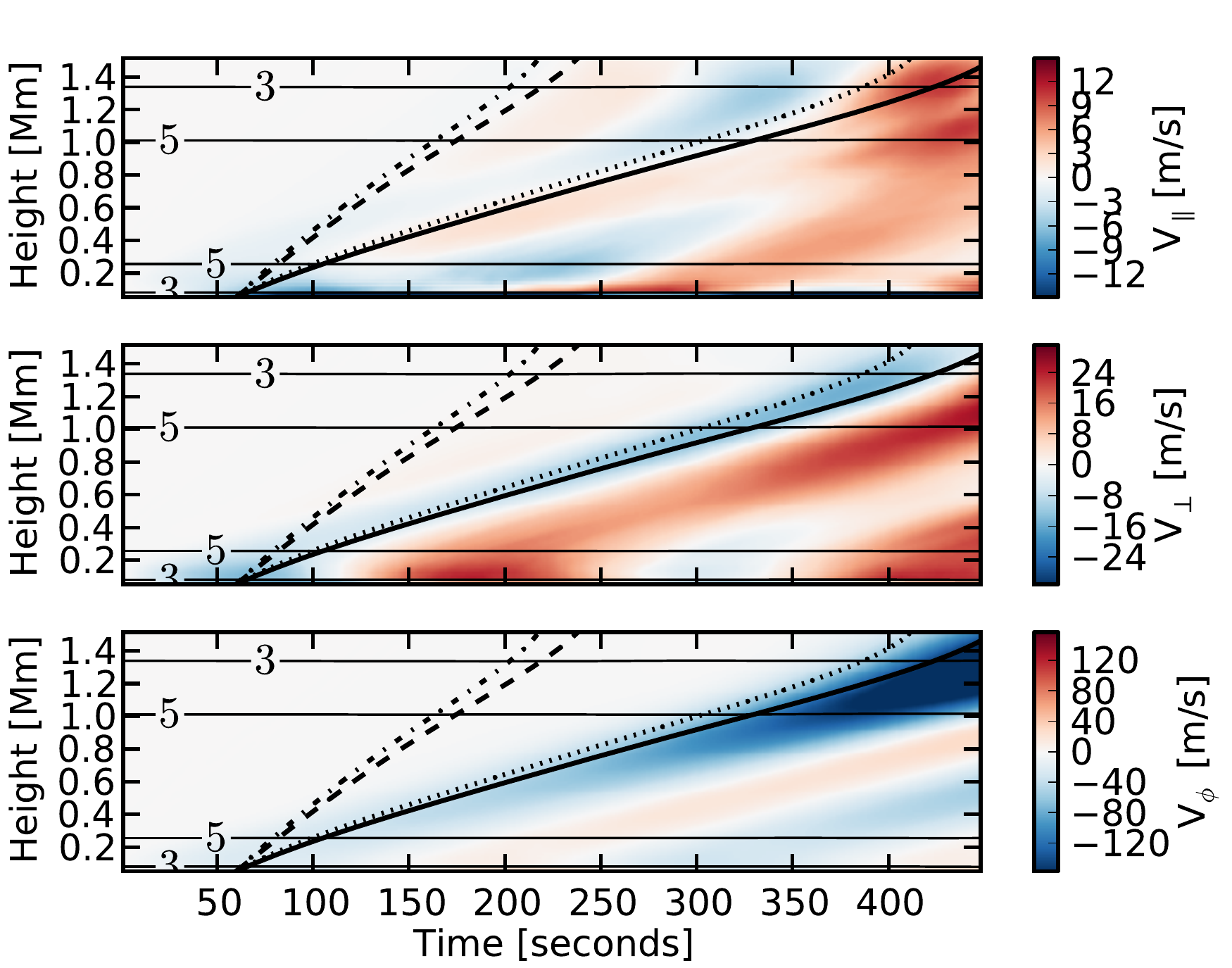}
		\caption{Archimedean Spiral Type Driver}
		\label{fig:All_TD_wave_30:Sarch}
	\end{subfigure}
	
	\begin{subfigure}[b]{0.49\textwidth}
		\includegraphics[width=\columnwidth]{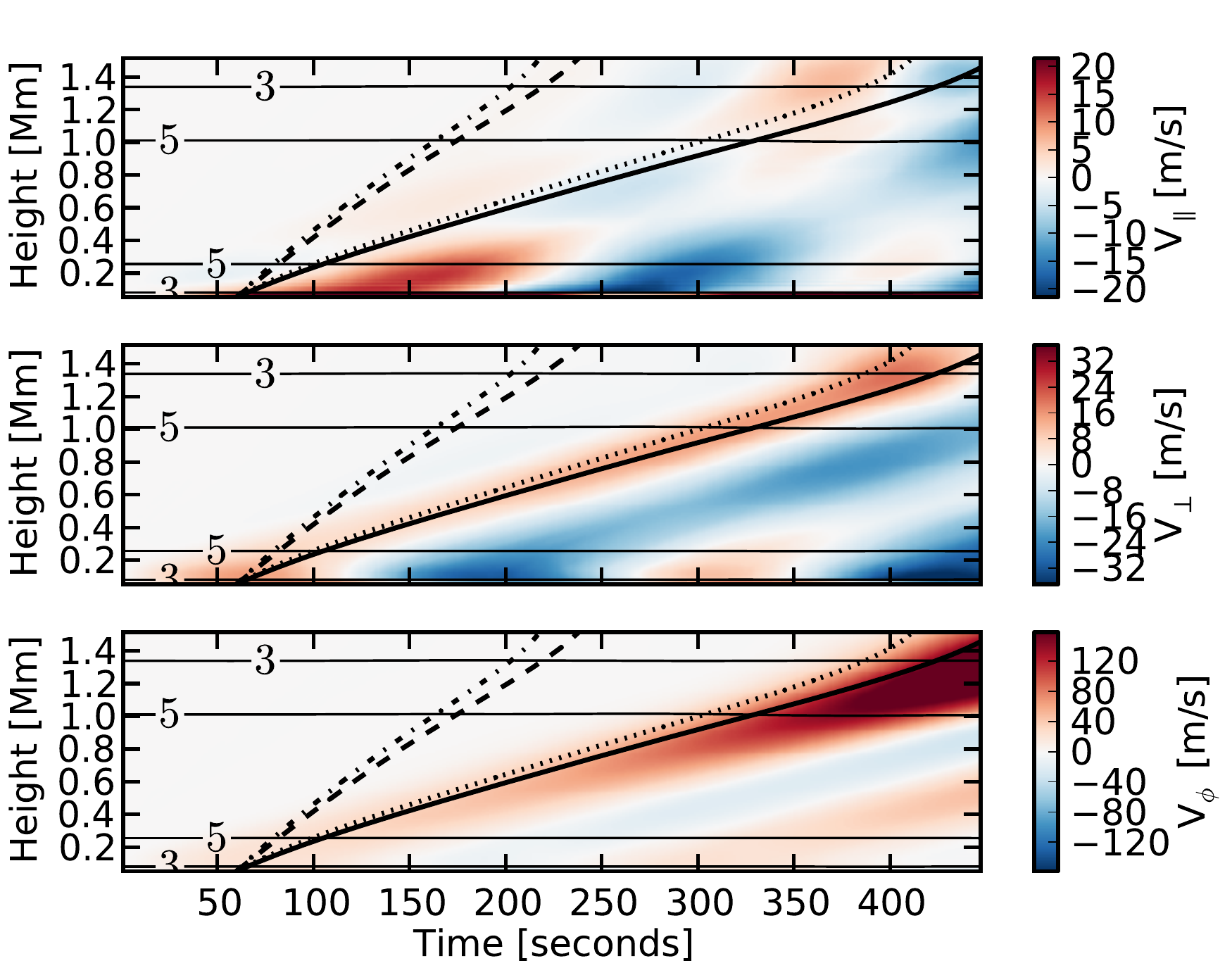}
		\caption{Logarithmic Spiral Type Driver}
		\label{fig:All_TD_wave_30:Slog}
	\end{subfigure}
	\caption{Decomposed velocity perturbation time-distance diagrams along the flux surface at radius $r = 468$ km (approximately central in the magnetic flux tube) for all simulated drivers. Horizontal black lines are plasma-$\beta$ contours, over-plotted are characteristic background speeds, the dot-dashed line is the fast speed ($v_f$), the dashed line is the sound speed ($c_s$), the dotted line is the Alfv\'en speed ($v_A$) and the solid line is the slow speed ($v_t$).}
	\label{fig:All_TD_wave_30}
\end{figure*}

To identify the observed MHD wave modes we shall initially consider the phase speed of the perturbations in the time-distance diagrams. 
 In our numerical domain, both inside and outside the flux tube, there is plasma $\beta > 1$.
 In analysing the results we identified the fast sausage mode, the slow kink mode and the Alfv\'en mode.
 To aid in the analysis of Fig. \ref{fig:All_TD_wave_30} overplotted are the Alfv\'en speed $v_A$ and sound speed $c_s$, as well as the speed of the fast magneto-acoustic wave (fast speed) $v_f^2 = \sqrt{c_s^2 + v_A^2}$ and the slow speed (or tube speed) $v_t^{-2} = \sqrt{c_s^{-2} + v_A^{-2}}$ for the equilibrium background, starting at $60$ s, the first peak of the driver amplitude. 
 It should be noted that this analysis is still an approximation of our simulated system because we have non-constant, non-uniform, non-straight magnetic field in a stratified solar atmosphere, where one would expect the observed phase speed to deviate from these first-order estimations, as can be seen in Fig. \ref{fig:All_TD_wave_30}.

First, we take the case of the horizontal driver, Fig. \ref{fig:All_TD_wave_30:horiz}, in the most detail. 
 In the $V_\parallel$ component we expect to see the fast sausage mode being the dominant mode, which is observed.
 There is also a weaker presence of a perturbation with the phase speed closer to that of the Alfv\'en and slow speeds but offset from the starting point of the over-plotted lines; This is attributed to the coupling of the wave modes in our non-homogeneous plasma.
 In the $V_\perp$ component the presence of a slow kink mode travelling close to the tube speed $v_t$ (solid line).
 This mode is the dominant contribution in this panel and is approximately two times stronger when compared the perturbations in the parallel component. 
 Finally, the azimuthal velocity component ($V_\phi$) has a very small contribution, of an order of magnitude less, travelling at the Alfv\'en speed, which we attribute to our driver not being perfectly centred upon the flux tube axis.

Comparing the results of the wave excitation by the vertical periodic driver to that of the horizontal driver, it is easy to draw parallels in the description.
 However, there are some key differences. 
 In this case, of wave excitation by the vertical driver, most of the perturbation is in the $V_\parallel$ velocity component, with a much stronger contribution from the fast sausage mode ($\approx 20 \times$ stronger than $V_\perp$).
 There is also evidence of a rapidly spatially damped mode observed in the top panel of Fig. \ref{fig:All_TD_wave_30:vert}.
 This spatial damping is attributed to the expansion of the magnetic flux tube, and the dispersion of the wave energy over a wider volume as the tube expands.
 The $V_\perp$ component on the vertical driver's time-distance diagram is very weak, with only a weak fast kink mode component easily visible, apart from some small reflection from the top boundary. 
 Finally, the vertical driver's $V_\phi$ component is, like its horizontally driven $V_\phi$ counterpart, substantially weaker than the other two components.

Next, we analyse the results of the three simulations with torsional drivers.
 The time-distance diagrams for the three different torsional drivers have similar properties; the vast majority of the perturbation for all the torsional drivers is, as expected, in the $V_\phi$ component.
 The other two components are of an order of magnitude less than the values of $V_\phi$.
 The time-distance diagrams for the uniform torsional and the Archimedean spiral driver, Figs. \ref{fig:All_TD_wave_30:Suni} \& \ref{fig:All_TD_wave_30:Sarch}, have in their $V_\parallel$ component evidence of both the fast sausage mode travelling close to the fast speed, and another very weak mode travelling close to the slow speed.
 We attribute this to the same wave mode coupling as observed in the horizontal driver's time-distance diagram. 
 The logarithmic spiral simulation has a more predominant signature in the $V_\parallel$ velocity component, where the rapidly spatially damped slow sausage mode is the predominant signal, similar to that observed in the case of the vertical driver. 
 In all three torsional drivers there is a notable presence of the coupled slow kink mode in the $V_\perp$ component. 

To gain a clearer understanding of the relative strength of each wave mode identified in Fig. \ref{fig:All_TD_wave_30} we now calculate the percentage wave energy flux carried by each component.

\subsection{Wave Energy Flux}\label{sec:energy_flux}

\begin{figure*}
	\centering
	\begin{subfigure}[b]{0.49\textwidth}
		\includegraphics[width=\columnwidth]{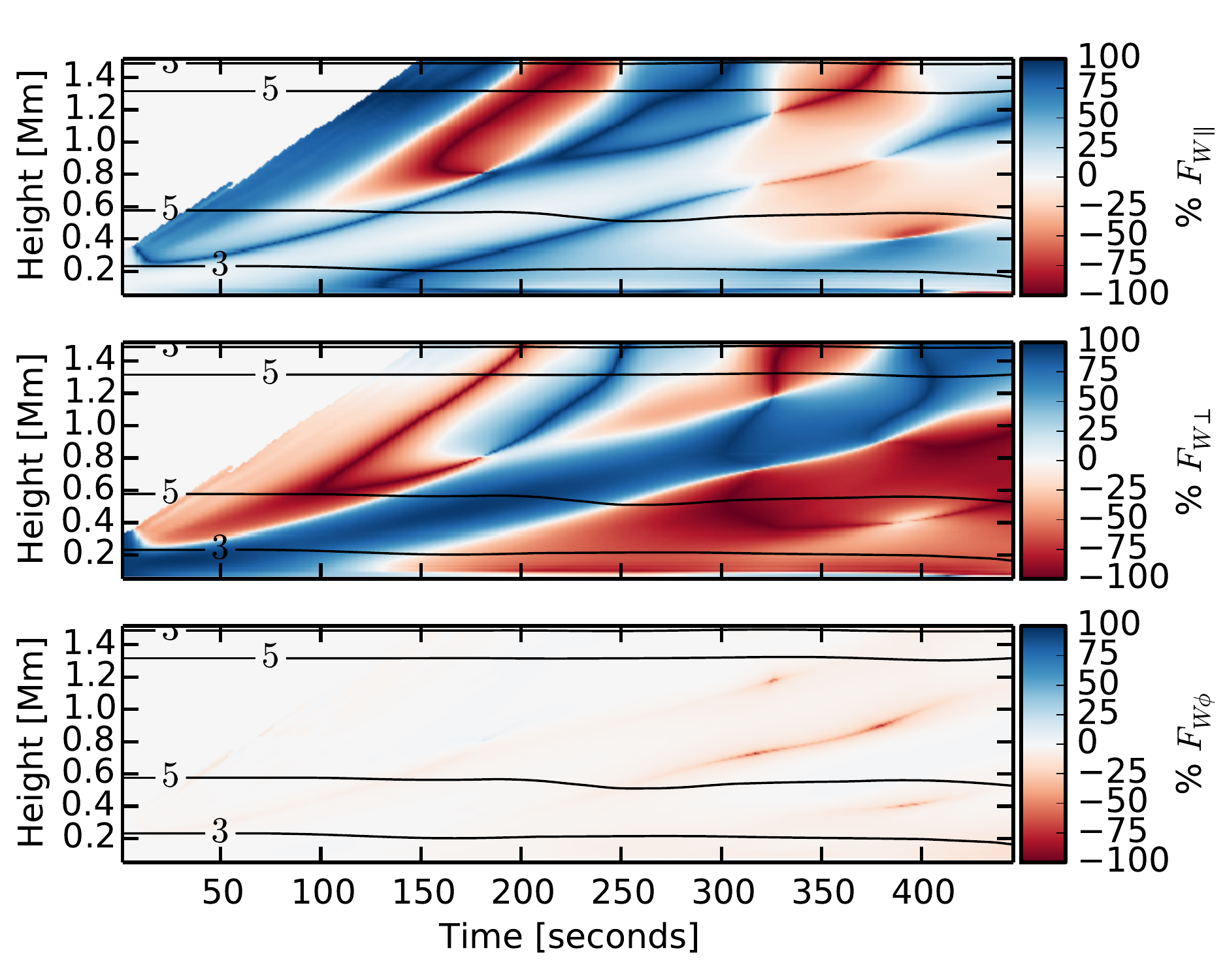}
		\caption{Horizontal Driver}
	\end{subfigure}
	\begin{subfigure}[b]{0.49\textwidth}
		\includegraphics[width=\columnwidth]{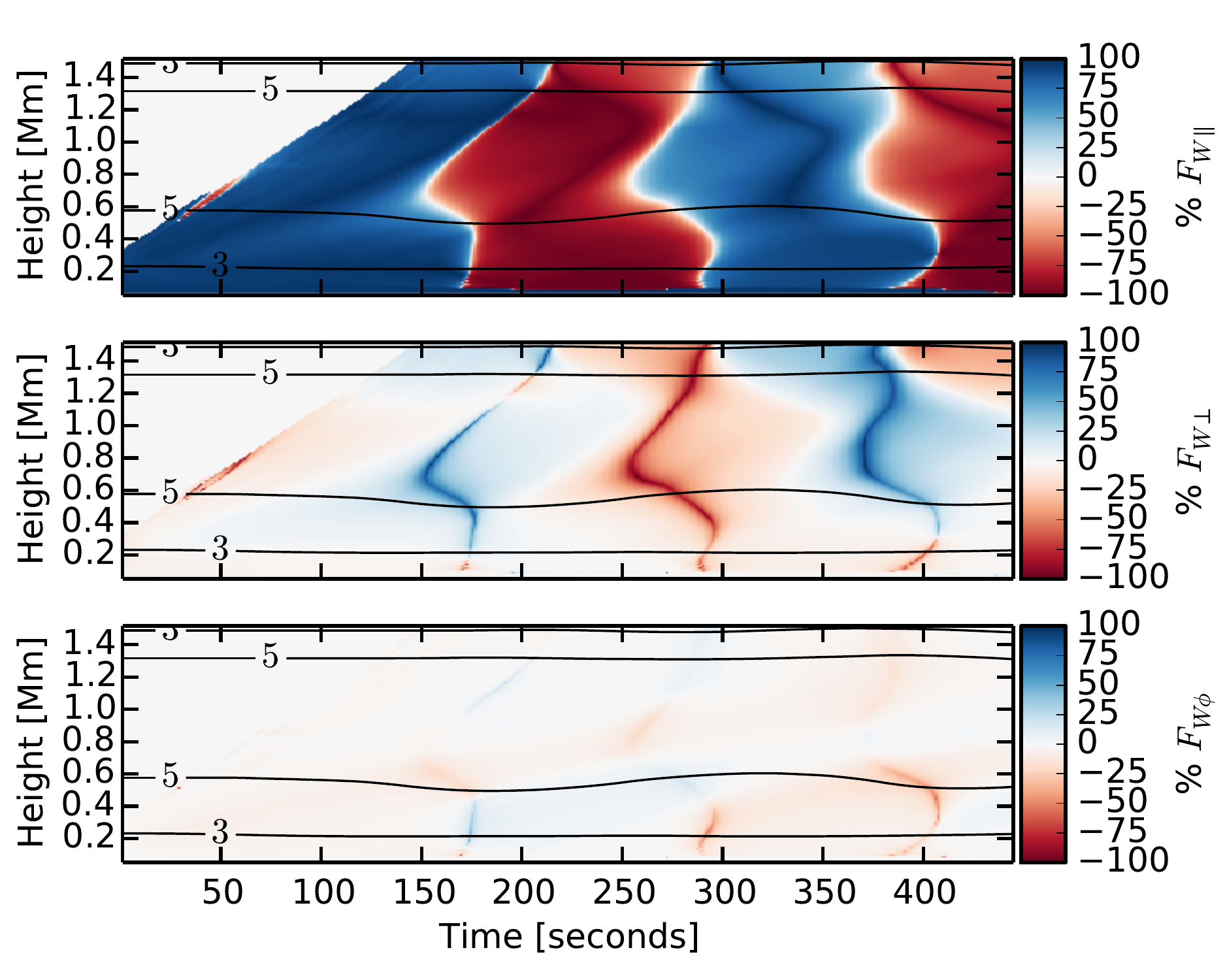}
		\caption{Vertical Driver}
	\end{subfigure}
	
	\begin{subfigure}[b]{0.49\textwidth}
		\includegraphics[width=\columnwidth]{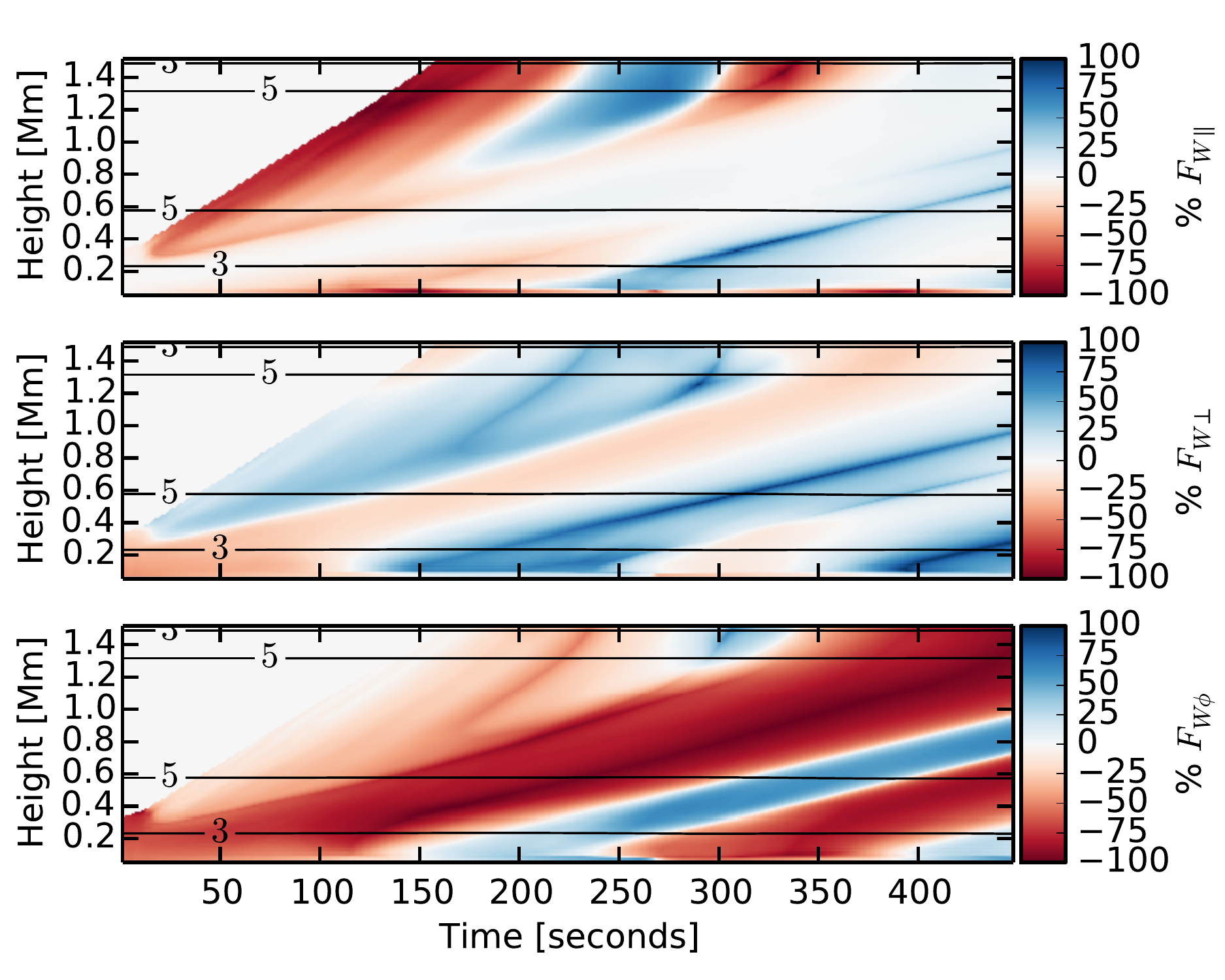}
		\caption{Uniform Torsional Driver}
	\end{subfigure}
	\begin{subfigure}[b]{0.49\textwidth}
			\includegraphics[width=\columnwidth]{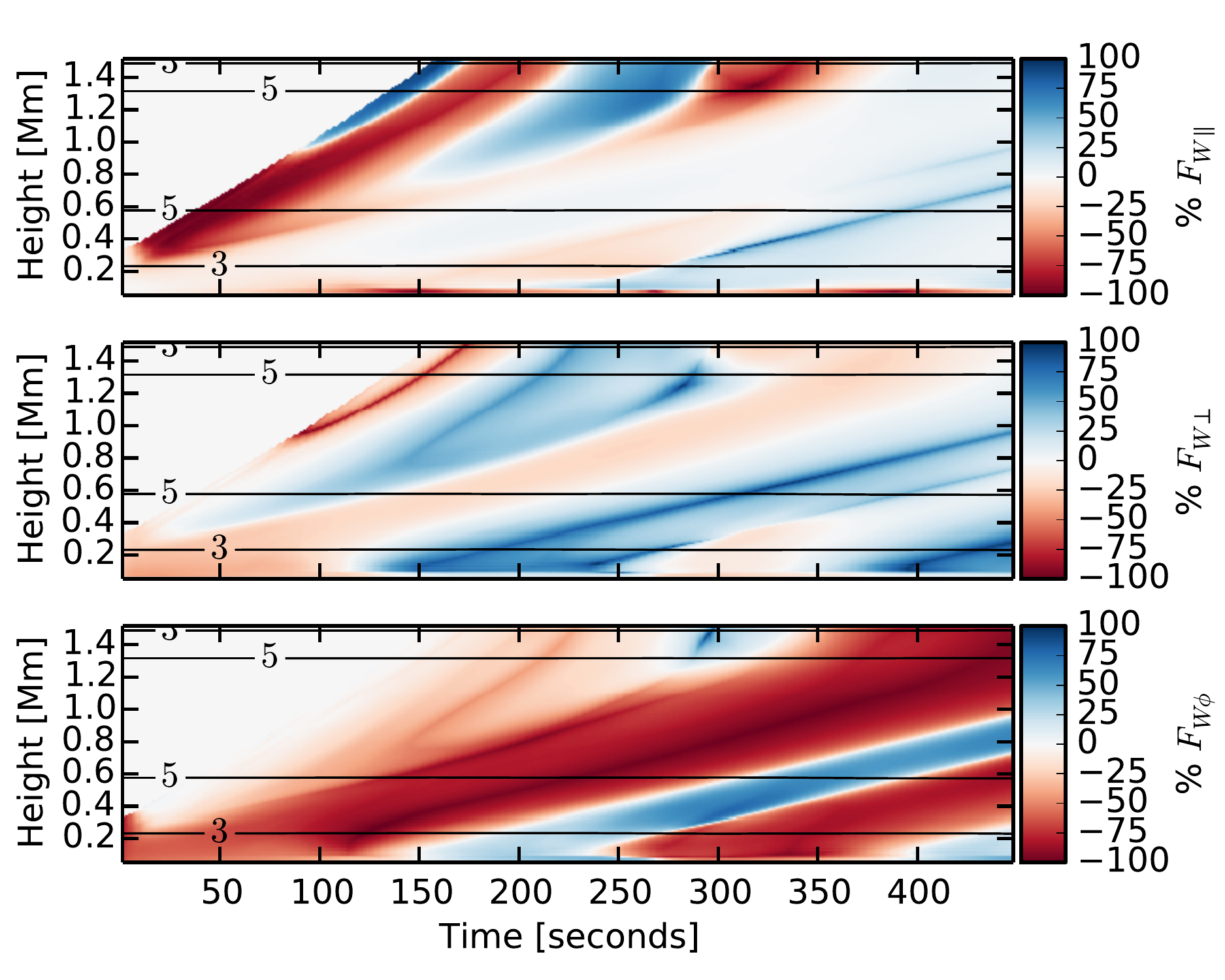}
			\caption{Archimedean Spiral Type Driver}
	\end{subfigure}
	
	\begin{subfigure}[b]{0.49\textwidth}
		\includegraphics[width=\columnwidth]{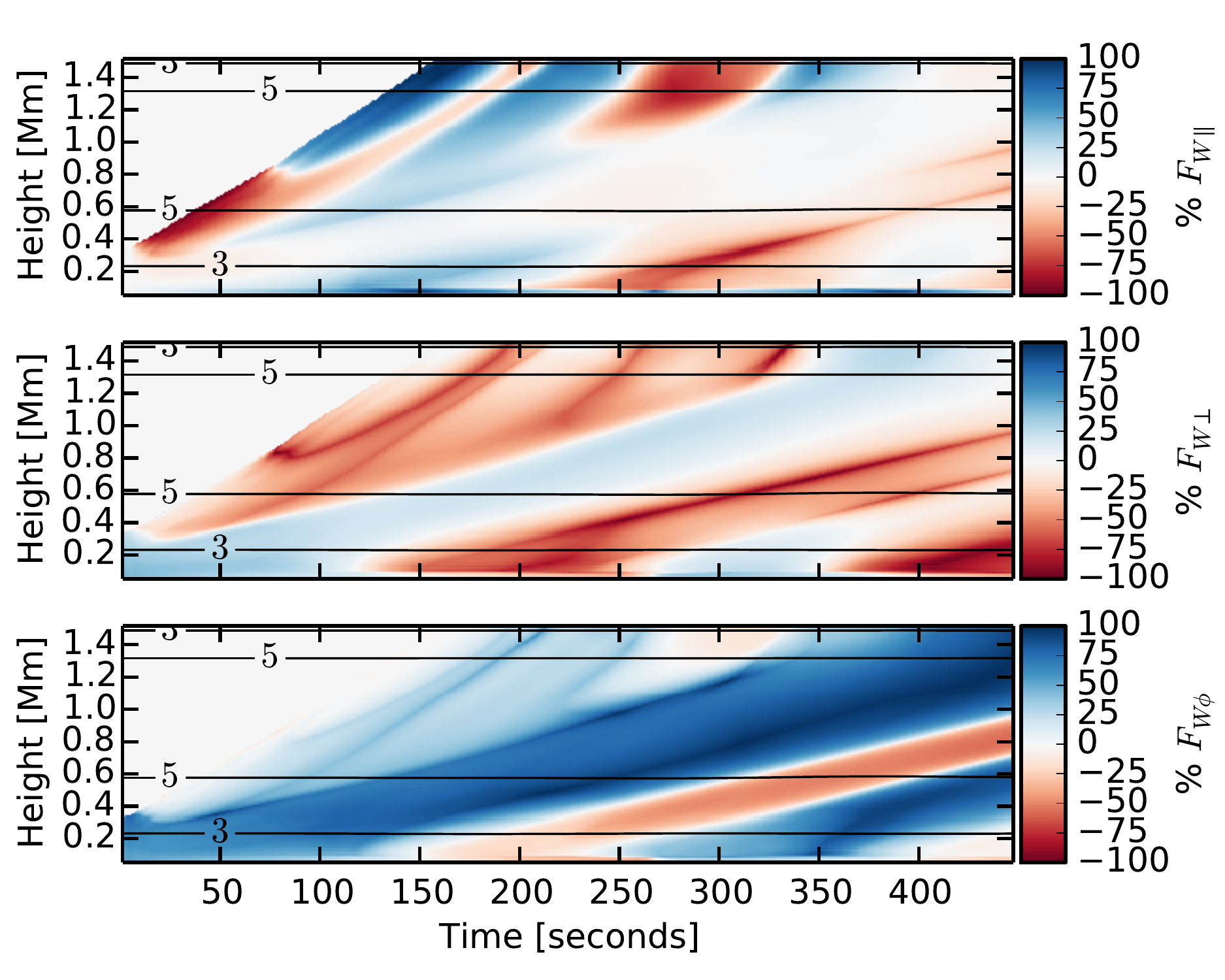}
		\caption{Logarithmic Spiral Type Driver}
	\end{subfigure}
	\caption{Decomposed wave energy flux time-distance diagrams along the flux surface at radius $r = 468$ km (approximately central in the flux tube) for all simulated drivers. The three components of Energy flux ($F_\parallel$, $F_\perp$ and $F_\phi$) are calculated, then, the proportion for each component is shown for a strip up the flux surface.}
	\label{fig:All_Flux_percent_TD}
\end{figure*}

To calculate the relative strengths of the excited waves we compute the `wave energy flux' vector everywhere in the domain using Equation \ref{eq:wave_energy}.
\begin{equation}
	\vec{F}_{wave} \equiv \widetilde{p}_k \vec{v} + \frac{1}{\mu_0} \left(\vec{B}_b \cdot \vec{\widetilde{B}}\right) \vec{v} - \frac{1}{\mu_0}\left(\vec{v} \cdot \vec{\widetilde{B}} \right) \vec{B}_b,
	\label{eq:wave_energy}
\end{equation}
where a subscript $b$ represents a background variable, a tilda represents a perturbation from the background conditions and $p_k$ represents kinetic pressure.

\begin{figure}
	\includegraphics[width=\columnwidth]{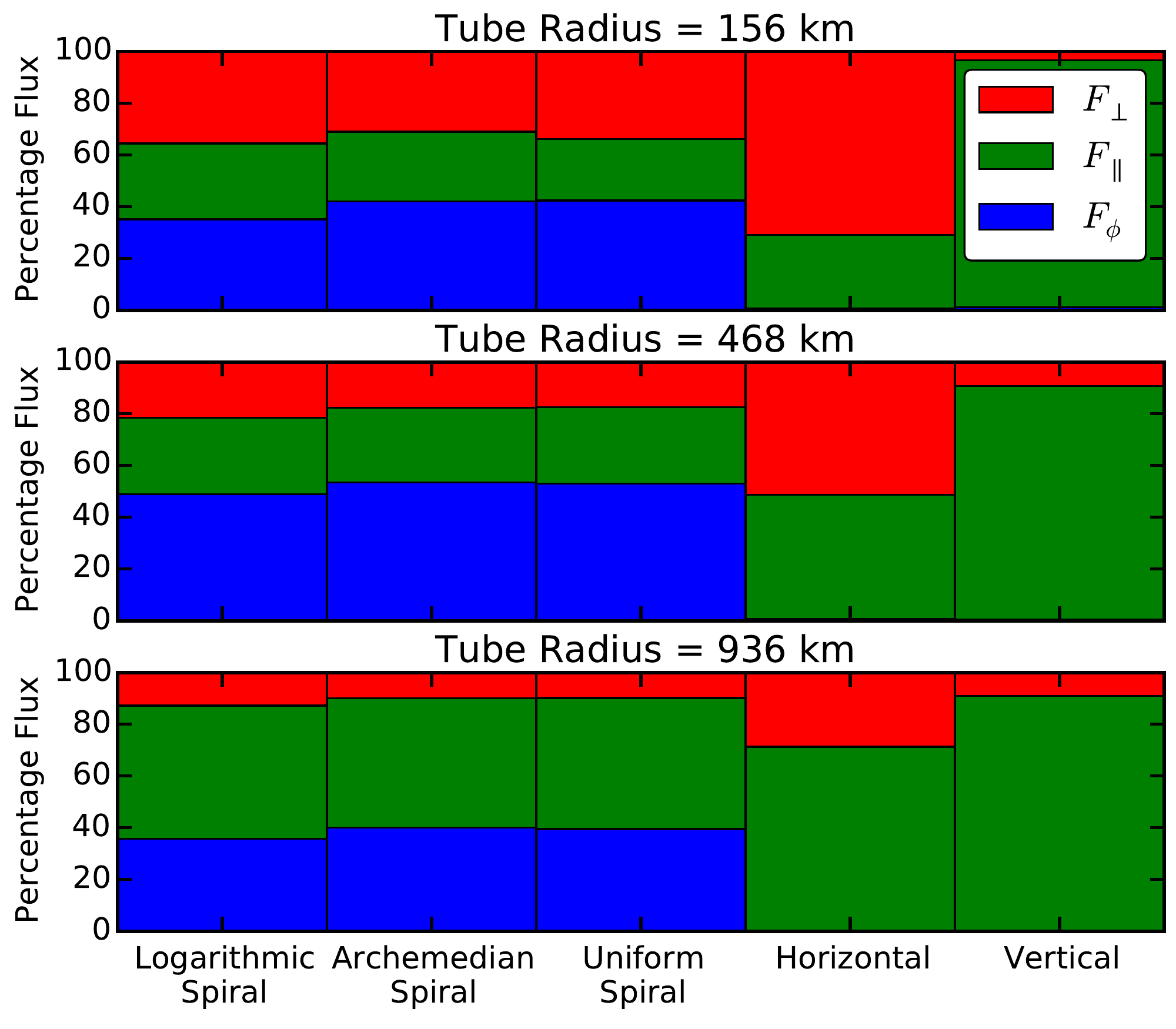}
	\caption{Percentage total available energy flux comparison (calculated using Equations \ref{eq:flux_par} - \ref{eq:flux_phi}), for all drivers and all flux surfaces. The $F_\parallel$ component is shown as green, the $F_\perp$ component is shown in red and the $F_\phi$ component is shown in blue.}
	\label{fig:flux_bar_graph}
\end{figure}

This equation has been widely used to calculate the energy contained in linear MHD perturbations.
 It is discussed in detail in \cite{bogdan2003} where it is compared to the `true' MHD flux for linear perturbations and found to be generally clearer. 
 It is used in \cite{vigeesh2009, vigeesh2012, khomenko2012}. 
 For a full derivation and discussion relating to time-averaging see \cite{leroy1985}.
 Calculating wave energy flux using Equation \ref{eq:wave_energy} provides a vector which is useful in plotting time distance diagrams and analysing wave modes.
 However, when averaging it to compute the values used in Figure \ref{fig:flux_bar_graph} the nature of the wave energy flux means that if there are standing waves in the domain the fluxes would cancel and therefore under represent the amount of energy excited into that component.
 While it is not expected to find standing waves in these simulations, we use Equations \ref{eq:flux_par} - \ref{eq:flux_phi} below, from \citet{vigeesh2012} and \citet{khomenko2012} to calculate the average `available' energy flux.
 These equations give an estimate of the `available' flux as energy density multiplied by wave speed.
 Which is advantageous for calculating the total average as it results in a positive value that is not negated by standing waves.
 
 \begin{align}
 	F_\parallel = \rho v_\parallel^2 c_s,\label{eq:flux_par}\\
 	F_\perp = \rho v_\perp^2 v_A,\label{eq:flux_perp}\\
 	F_\phi = \rho v_\phi^2v_A.\label{eq:flux_phi}
 \end{align}
 here, $F_\parallel$, $F_\perp$ and $F_\phi$ are the parallel, perpendicular and azimuthal components of energy flux respectively.

Once the wave energy flux has been computed, it is decomposed into parallel, perpendicular and azimuthal components using the same method as the velocity vector. 
 Using the analysis method outlined in Section \ref{sec:3d_analysis} time-distance diagrams are computed for the percentage wave energy flux (see Fig. \ref{fig:All_Flux_percent_TD}). 
 The percentage values are plotted to highlight the relative strengths of the excited wave modes, and to enable a comparison of which modes are dominant. 
 The absolute average energy flux over all heights is summed for all times for each component.

By comparing Figs. \ref{fig:All_TD_wave_30} \& \ref{fig:All_Flux_percent_TD} we find that for the wave modes excited by the horizontal driver $60$\% of the energy flux is in the perpendicular component $F_\perp$ which is attributed to the slow kink mode.
 The rest of the flux is in the parallel component $F_\parallel$. 
 The vertical driver simulation has $79.3$\% of the energy flux in the $F_\parallel$ component, identified as the fast sausage mode, with the $F_\perp$ component only contributing $12.5$\%. 
 The simulations with spiral drivers all have up to $60$\% of their energy flux in the azimuthal component $F_\phi$. 
 The logarithmic spiral source excites a slightly higher percentage of the flux in the slow kink mode and the fast sausage mode, in comparison to the uniform torsional and Archimedean spiral driver.

The summarised energy flux results, and their equivalents for different flux tube radii are shown in Fig. \ref{fig:flux_bar_graph}.
 With reasonable accuracy we can attribute each of the energy flux components shown to one or two MHD wave modes.
 The $F_\parallel$ component is generally the fast sausage mode. 
 The $F_\perp$ component is almost exclusively excited by the slow kink mode.
 Finally, the $F_\phi$ is attributed to the Alfv\'en mode.
 Another interesting result is that the type of spiral driver used has a minimal impact upon the amount of flux in each wave mode (see Fig. \ref{fig:flux_bar_graph}).
 This could be dependent upon the spiral expansion factor used in the logarithmic and Archimedean spirals, which could be the subject of a further parameter study.

\subsection{Flux Tube Radius}
The plasma properties vary within the computational domain due to the magnetic field configuration.
 This also means that the wave propagation on the surface of a flux tube is dependent upon its radius. 
 We define the radius of the flux tube at the top of the domain and as its initial radius.
 There are an arbitrary number of definable flux tube surfaces in our domain as defined from the top outer edge of the domain inwards. 
 To demonstrate the difference in propagation caused by the change in plasma properties, especially $\beta$, with a change in radii we have computed all the analysis for three different flux tubes, with radii of $r=936$ km, $r=468$ km and  $r=156$ km; These radii are chosen to represent a good spectrum across the domain.

The results of the flux calculations are summarised in Fig. \ref{fig:flux_bar_graph}.
 The smallest radius flux tube, shown in the top panel, shows that, for the torsional driver simulations, less azimuthal ($F_\phi$) flux is generated closer to the axis of the flux tube. 
 This is expected due to a higher magnetic pressure towards the axis of the tube; the flux is, instead, excited evenly in the parallel ($F_\parallel$) and perpendicular ($F_\perp$) components as predominately kink and sausage modes. 
 For higher radii surfaces the $F_\parallel$ component dominates the $F_\perp$ component; as the distance from the axis increases the influence of the kink mode decreases.
 In the case of the horizontal and vertical drivers, most of the flux is excited in the slow kink and sausage modes respectively.
 In the horizontal case, for the larger radius tube, the sausage mode, in the $F_\parallel$ component, again begins to dominate the kink mode, in the $F_\perp$ component.

\section{Conclusion}\label{sec:conclusion}
In this paper we have presented 3D numerical simulations showing wave propagation from simulated photospheric drivers, up through the low solar atmosphere towards the transition region.
 A novel, practical method for decomposing the velocity perturbations into a parallel, perpendicular and azimuthal components in a three dimensional geometry was developed using field lines to trace a volume of constant flux and then creating a set of polygons along the surface of the volume from which a perpendicular vector could be computed. 
 This method was then employed to identify the excited wave modes propagating upwards from the photosphere and to compute the percentage energy contribution of each mode.

Simulations were run mimicking five types of photospheric motions: horizontal and vertical drivers, and uniform torsional, Archimedean and logarithmic spiral velocity fields were modelled.
 The resulting perturbations were analysed, the wave modes identified and their percentage wave energy flux contribution determined. 
 We have determined that for all drivers with a torsional component the main contribution to the flux was the Alfv\'en wave.
 While the vertical driver mainly excites the fast- and slow-sausage modes and the horizontal driver primarily generated the slow kink mode.
 Further extensions to this work will include varying period of the drivers and extending the vertical extent of the atmosphere to incorporate the low corona.
 With this larger atmosphere we may be able to determine the energy transport through the domain, where mode conversion is an important ingredient.
 
There have been many recent observations of torsional motions recently observed in the solar photosphere, at the very limits of modern telescopes.
 Modelling these motions which have the potential to be ubiquitous in the photosphere, demonstrates the potential for large amounts of Alfv\'e{n} wave excitation in small scale magnetic structures.
 Also shown via the numerical simulations in this work is the damping and propagation properties of the excited MHD waves at various heights in a realistic expanding magnetic flux tube.
 Further extensions of this work into the transition region and corona would allow more, indirect, comparison to observational results of heating in the transition region.

\section*{Acknowledgments}
The authors would like to acknowledge the NumPy, SciPy \citep{jones2001}, Matplotlib \citep{hunter2007} and MayaVi2 \citep{ramachandran2011} Python projects for providing the computational tools to analyse the data. SM would like to thank Peter Wesson for helpful discussion regarding spiral vector fields. RE is thankful to the Science and Technology Facilities Council (STFC) and NSF, Hungary (OTKA, Ref. No. K83133) and acknowledges M. K\'eray for patient encouragement.

\bibliographystyle{apj}
\bibliography{smumford_etal_2012}{}

\appendix
\section{Cutoff Frequency}

Due to the gravitationally and magnetically stratified atmosphere used in these simulations a cutoff frequency will be introduced for all the MHD wave modes propagating in the domain.
While the results presented in Section \ref{sec:results} indicate the existence of many different propagating wave modes travelling along the field lines with different phase speeds, it is worth discussing the potential effects of the cutoff frequencies on these results.

Much analytical work has been done to determine the cutoff frequencies for different wave modes in a variety of increasingly complex models.
\cite{defouw1976} studies longitudinal waves in the thin-flux tube approximation and \cite{roberts2006} considers the cutoff frequency of the slow mode in a stratified atmosphere with an embedded uniform magnetic field.
\cite{spruit1981} also considers the case of a thin flux tube, embedded in a non-magnetic background and then calculates the cutoff frequency for transverse wave modes.
However, more recent work \citep{dymova2005, ruderman2008} and \citep[private communication]{andries2014} calls this result into question.
\cite{musielak2007} studied torsional wave modes in a thin flux tube and found no evidence of a cutoff frequency for the Alfv\'en mode.

\begin{figure*}[h]
	\centering
	\begin{subfigure}[b]{0.49\textwidth}
		\includegraphics[width=\columnwidth]{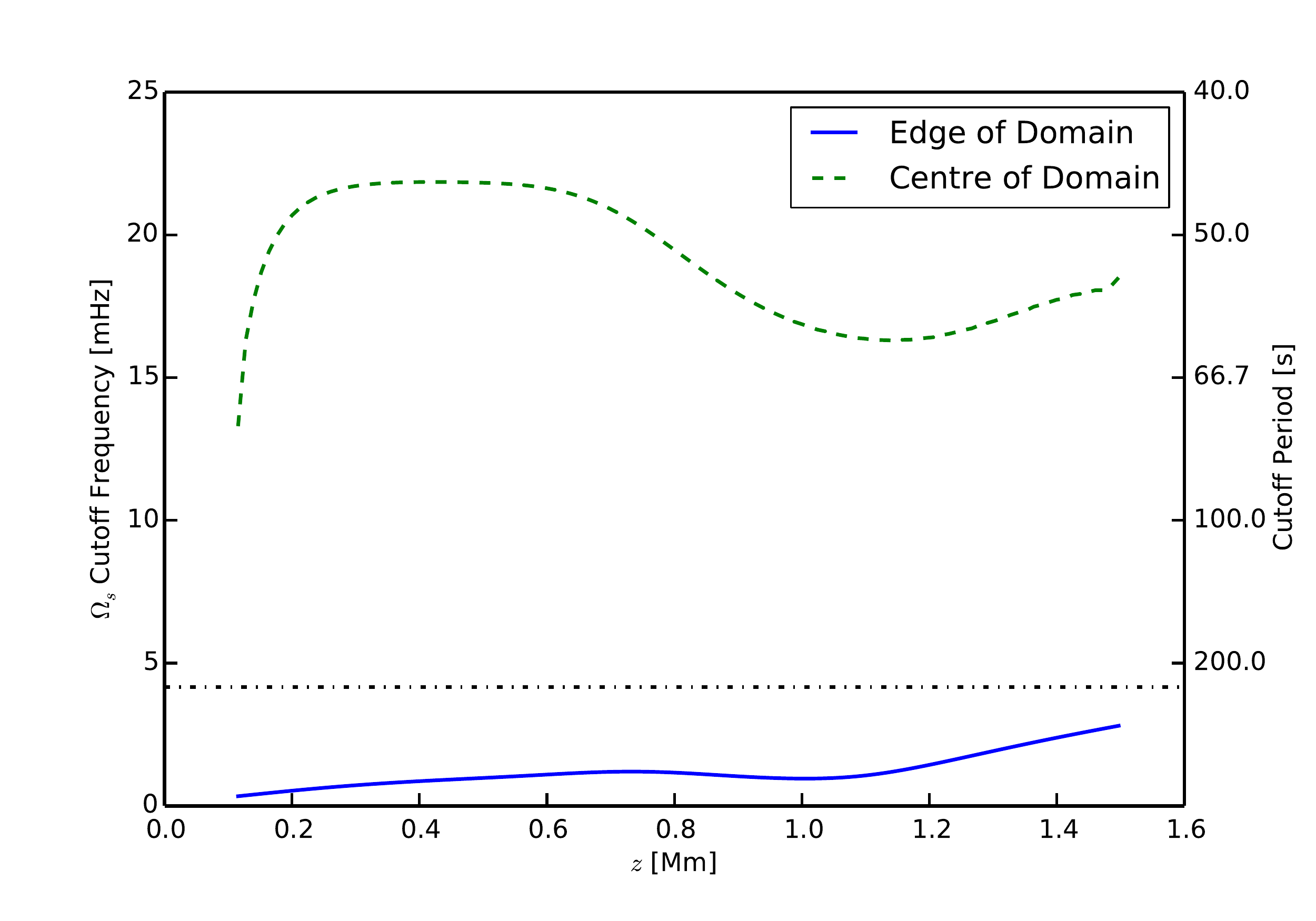}
		\caption{}
		\label{fig:slow_cutoff}
	\end{subfigure}
	\begin{subfigure}[b]{0.49\textwidth}
		\includegraphics[width=\columnwidth]{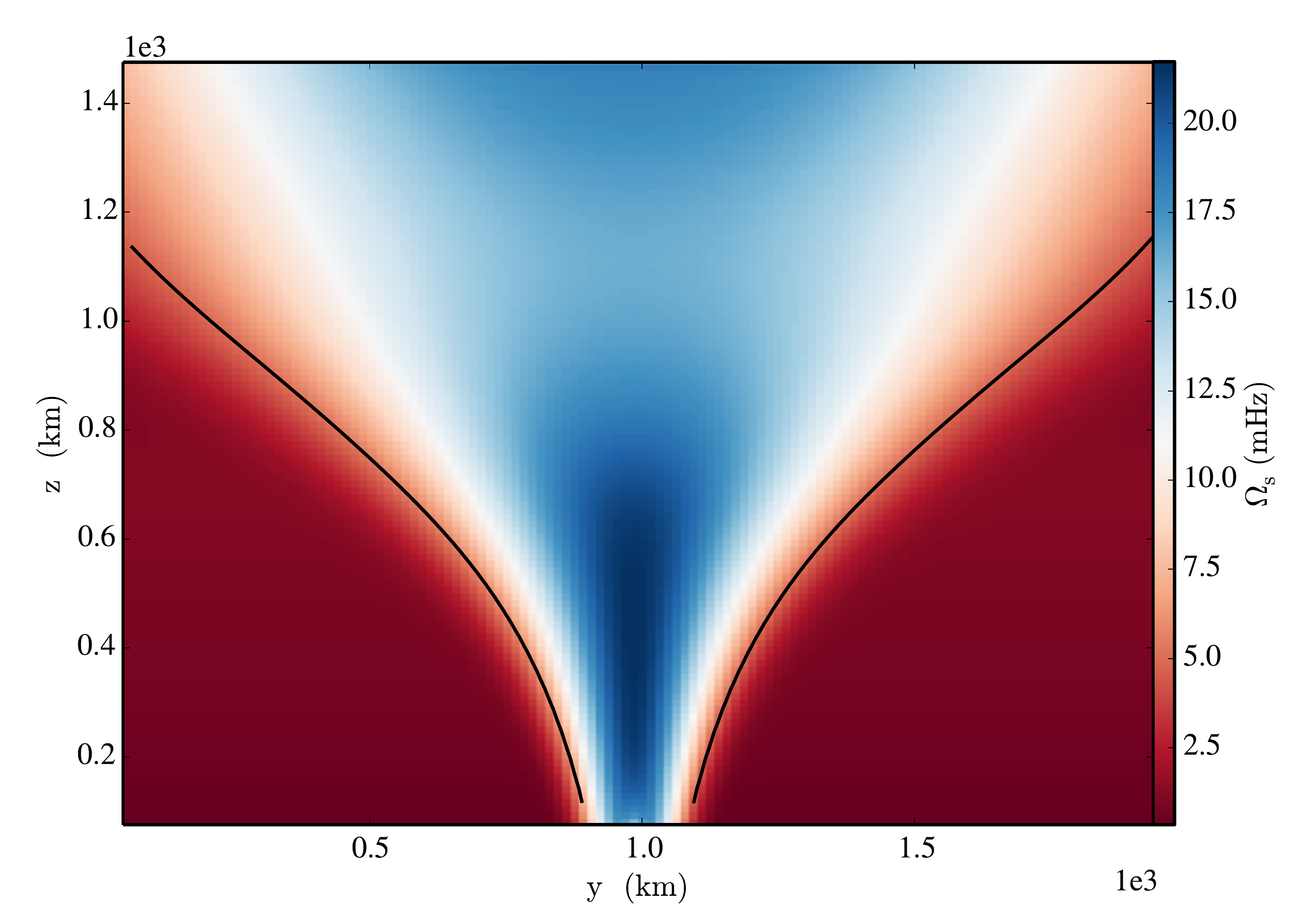}
		\caption{
		}
		\label{fig:slow_cutoff_slice}
	\end{subfigure}
	\caption{(a) The slow magnetoacoustic mode cutoff frequency for the simulation domain as derived in \cite{roberts2006}.
			The blue solid line show the cutoff frequency at the edge of the domain, in the weakly magnetic background.
			The green dashed line shows the cutoff frequency in the centre of the magnetic flux tube, where the magnetic field is strongest.
			The black dash-dotted line shows the $240$ s period used for the drivers in this work.\\
			(b) A slice through the simulation domain along the centre of the domain in the x direction.
			The black contour traces the frequency of the driver used in this work ($4.1$ mHz).}
	\label{fig:cutoff}
\end{figure*}

To further the discussion of cutoff frequencies in this case of a expanding flux tube in a magnetic stratified background, the slow mode cutoff from \cite{roberts2006} was calculated for all points in the simulation domain.
These results are shown in Figure \ref{fig:cutoff}.

It can be seen that the cutoff frequency calculated above prevents a slow mode with the driver period of $240$ s from propagating in most of the domain.
This is not seen as an issue, firstly, due to the deviation of the analytical model from the simulation configuration and, secondly, due to the nature of the slow mode perturbations observed in these simulations. 
The only slow mode observed is the kink mode for the horizontal driver (shown in the second panel of Figure \ref{fig:All_TD_wave_30:horiz}).
This mode is sufficiently different from the slow mode considered by \cite{roberts2006}, and considering recent work on cutoff frequencies of transverse waves, the analytical interpretation of its cutoff frequency is complex.

In conclusion, from the results presented in Figures \ref{fig:All_TD_wave_30} \& \ref{fig:All_Flux_percent_TD} it is clear that these simulations generate propagating wave modes.
Due to the complexity of these simulations it is difficult to separate and identify the effects of the different linear MHD wave modes on the plasma properties.
This is also true for the cutoff frequency calculations, where it is clear that the analytic theory also does not satisfactorily describe the simulation domain used in this work.

\end{document}